\documentclass[pra,aps,amsmath,amssymb,floatfix,11pt,superscriptaddress]{revtex4-2}
\usepackage{graphicx,comment}% Include figure files
\usepackage{dcolumn}% Align table columns on decimal point
\usepackage{bm,bbm}% bold math
\usepackage{soul}
\usepackage{yfonts}
\usepackage{amsmath}
\usepackage{float}
\usepackage{xcolor}
\usepackage[colorlinks=true,linkcolor=blue,citecolor=blue,urlcolor=blue]{hyperref}
\def\bmlambda{\boldsymbol{\lambda}}
\def\Tr{\mbox{Tr}}
\def\sep{{\rm sep}}
\def\Q{{\boldsymbol{Q}}}
\def\F{{\boldsymbol{F}}}
\def\W{{\boldsymbol{W}}}
\def\U{{\boldsymbol{U}}}
\def\X{{\boldsymbol{X}}}
\def\L{{\hat{L}}}
\def\hG{{\hat{G}}}
\def\hU{{\hat{U}}}
\def\hV{{\hat{V}}}
\begin{document}
\title{Mitigating sloppiness in joint estimation of successive squeezing parameters}
%%%5
\author{Priyanka~Sharma}
\email{priyankasharma@bhu.ac.in}
\affiliation{Department of Physics, Banaras Hindu University, Varanasi-221005, India}

\author{Stefano~Olivares}
\email{stefano.olivares@fisica.unimi.it}
\affiliation{Dipartimento di Fisica ``Aldo Pontremoli'', Università  di Milano, I-20133 Milano, Italy}

\author{Devendra~Kumar~Mishra}
\email{dkmishra.phys@bhu.ac.in}
\affiliation{Department of Physics, Banaras Hindu University, Varanasi-221005, India}

\author{Matteo G.~A.~Paris}
\email{matteo.paris@fisica.unimi.it}
\affiliation{Dipartimento di Fisica ``Aldo Pontremoli'', Università  di Milano, I-20133 Milano, Italy}

%%%
\begin{abstract}
When two successive squeezing operations with the same phase are applied to a field mode, reliably estimating the amplitude of each is impossible because the output state depends solely on their sum. In this case, the quantum statistical model becomes sloppy, and the quantum Fisher information matrix turns singular. However, estimation of both parameters becomes feasible if the quantum state is 
subjected to an appropriate scrambling operation between the two squeezing operations. In this 
work, we analyze in detail the effects of a phase-shift scrambling transformation, optimized to reduce sloppiness and maximize the overall estimation precision. We also compare the optimized precision bounds of joint estimation with those of stepwise estimation methods, finding that joint estimation retains an advantage despite the quantum noise induced by the residual parameter incompatibility. Finally, we analyze the precision achievable by general-dyne detection 
and find that it may approach the optimal precision in some regimes.
\end{abstract}
%%%
\maketitle
\section{\label{1} Introduction}
Quantum metrology aims to optimize measurement techniques using quantum resources to exceed 
classical precision limits \cite{degen2017quantum,montenegro2025quantum}. In doing this, the task 
is to find an effective encoding of the quantity of interest onto the states of a quantum system, 
and to design a suitably optimized read-out measurement to effectively extract the encoded 
information. Quantum estimation theory provides the tools to perform this optimization and to 
establish the ultimate limits on the precision of parameter estimation in the quantum domain \cite{1976Helstrom, Holevo,MGAP,doi:10.1142/S0219749909004839}, and it aims to identify potential advantages for classical protocols by leveraging quantum resources \cite{drummond2013quantum, PhysRevLett.55.2409, PhysRevLett.87.270404}.

When dealing with a single parameter, one calculates the symmetric logarithmic derivative (SLD) 
operator and use it to determine the quantum Cramér-Rao bound (QCRB) \cite{PhysRevLett.72.3439,doi:10.1142/S0219749909004839}. This bound represents the highest precision that can be achieved in estimating the parameter. It is directly related to the quantum Fisher Information (QFI), which measures how much useful information can be extracted by the best quantum measurement. To enhance precision we need to maximize 
the QFI by suitable choice of quantum probes \cite{SHARMA2025130459, rossi2015entangled, bina2018continuous, PRXQuantum.2.020308, SHARMA24OC}. In many cases, these quantum-enhanced setups can surpass the performance of any classical measurement strategy, achieving higher accuracy in parameter estimation \cite{SethLloyd}.

In multi-parameter estimation, the task is to estimate the multiple unknown parameters encoded in a system \cite{ALBARELLI2020126311,liu2020quantum}. This field has received much attention in recent years, ranging from its applications in quantum imaging and super-resolution microscopy \cite{PhysRevX.6.031033}, to quantum-enhanced sensing of magnetic and gravitational fields \cite{taylor2008high, PhysRevLett.116.030801}, and from simultaneous estimation of multiple phases in optical interferometry \cite{PhysRevLett.111.070403} to quantum networks \cite{PhysRevLett.120.080501}.

As a matter of fact,  multiparameter quantum bounds are not a trivial generalization of the single-parameter ones  \cite{1976Helstrom}, as it is the case in the classical scenario. In quantum systems, the measurements we use to estimate the unknown parameters are quantum observables, which may not always commute with each other \cite{PhysRevA.61.042312}. This means that the best measurement for estimating one parameter might not work well for another, which makes it hard to estimate all parameters simultaneously with the highest possible precision. Because of this, the commonly used precision limit in quantum metrology, the QCRB based on SLDs, is often not tight when estimating multiple parameters at once. This actually reveals the unique quantum features of the system, such as measurement incompatibility \cite{PhysRevA.94.052108} or quantum correlations \cite{Szczykulska03072016}. To address this issue, a more accurate bound known as the Holevo Cramér-Rao bound (HCRB) has been derived \cite{Holevo,suzuki2016explicit,PhysRevLett.123.200503}. The HCRB takes into account the incompatibility of quantum measurements, as well as the weight one may want to assign to each parameter,  and provides a tight limit on how precisely all parameters can be estimated together \cite{ALBARELLI2020126311}. Refined bounds and adaptive strategies have been also proposed to quantify and overcome the limitations due to incompatible observables in different applications \cite{Szczykulska03072016,Demkowicz20,PhysRevLett.123.200503,RubioPhysRevA.101.032114,Lee9919344,razavian2020quantumness,Bressanini_2024}.

In multiparameter models with inefficient encoding, measurement outcomes may depend only on specific parameter combinations rather than on each parameter individually. This often arises when the encoding is distributed across time or space, effectively creating a model where individual parameters represent values at distinct times or locations. If the encoding fails to allow  estimation of all parameters, the model is termed sloppy \cite{frigerio2024overcomingsloppinessenhancedmetrology, PhysRevLett.121.130503, PhysRevA.94.052108, Candeloro_2024}. Sloppiness manifests in the quantum Fisher information (QFI) matrix as singularity or near-singularity—where one or more eigenvalues vanish or become negligible—rendering some parameters unestimable. While this limits achievable precision, sloppiness can be mitigated by strategically ``opening" the encoding box and scrambling information. This approach was successfully demonstrated in \cite{frigerio2024overcomingsloppinessenhancedmetrology} to mitigate—and in some cases eliminate—sloppiness in a two-phase Mach-Zehnder interferometer.

Motivated by these developments, we examine a paradigmatic two-parameter estimation model 
involving two consecutive squeezing operations, each inducing a hyperbolic phase shift 
on an initial probe \cite{PhysRevA.50.801,PhysRevA.47.2339}. The model is inherently sloppy, but this sloppiness can be reduced or suppressed by introducing a scrambling transformation \cite{frigerio2024overcomingsloppinessenhancedmetrology,jadesc}, a phase shift,  between the two squeezings. By optimizing both the scrambling phase and the initial phase of the probe, we minimize sloppiness, which in turn maximizes estimation precision, even in the presence of residual incompatibility.
We validate these findings by comparing the optimized precision bounds of joint estimation with those of stepwise methods, demonstrating that joint estimation retains an advantage despite quantum noise from parameter incompatibility. Furthermore, we analyze general-dyne detection and show that, in certain regimes, its precision can approach the optimal level. 

The paper is structured as follows. In Section \ref{Methodology}, we review multiparameter quantum estimation theory. In Section \ref{Proposed model} we describe our estimation scheme, where 
two successive hyperbolic phase-shift (squeezing) are imposed to an initial probe state, 
as well as the scrambling technique to optimally tune and reduce the sloppiness of the model. 
Section \ref{separate measurement} describe the separate (stepwise) measurement of parameters. In Section \ref{s:phase:space} we discuss the phase-space description of system's dynamics, and 
in Section \ref{generaldyne} we analyze joint estimation of the hyperbolic phases by general-dyne measurement. Section \ref{Conclution} closes the paper with some concluding remarks.
%%%
\section{Multi-parameter quantum estimation}\label{Methodology}
Multi-parameter statistical models may depend only on some functions of the parameters that are fewer than the number of initial parameters themselves. For example, a multi-parameter estimation problem consists of $l$ real unknown parameters $\lambda_1,\lambda_2,....\lambda_{l}$. We aim to build a statistical model that estimates all these parameters and depends on the parameters that are fewer than the number of initial parameters themselves. Reducing the dependency of the number of parameters is called {\em sloppiness}. Such sloppy statistical models are characterized by a degenerate Fisher information matrix \cite{frigerio2024overcomingsloppinessenhancedmetrology, ALBARELLI2020126311}.

In the multi-parameter quantum estimation problem we consider a statistical model, $\hat{\rho}_{\bmlambda}$, which is a family of density operators parameterized by a column vector of $l$ real unknown parameters $\bmlambda =(\lambda_1,\lambda_2,....\lambda_{l})^\intercal$. To estimate the parameters, we define a  positive-operator valued measure (POVM)  $\left\{ \hat{\Pi}_{\kappa} \right\}_{\kappa=1\,\ldots,l}$ such that $\hat{\Pi}_{\kappa}\geq 0$ and $\sum_{\kappa}\hat{\Pi}_{\kappa} = \mathbbm{1}$. The conditional probability for measurement outcome $\kappa$ is evaluated using Born, namely $p(\kappa|\bmlambda) = \text{Tr}[\hat{\rho}_{\bmlambda}\hat{\Pi}_{\kappa}]$. If we repeat the measurement $N$ times the set of outcomes, ${\rm K} =\{\kappa_1,\kappa_2,....\kappa_{N}\}$ (assumed to be identically distributed and independent), can be estimated using an estimator $\Tilde{\bmlambda}$. This estimator provides the value map from the space of measurement outcomes to the space of the possible values of the parameters $\bmlambda \in \mathbb{R}^l$, whose accuracy is generally addressed in terms of its mean-square error matrix 
\begin{equation}
   \mathcal{O} (\bmlambda) = \sum_{\kappa=1}^{l}p(\kappa|\bmlambda)(\Tilde{\bmlambda}-\bmlambda)^\intercal (\Tilde{\bmlambda}-\bmlambda).
\end{equation}
For locally unbiased estimators, the following conditions are satisfied
\begin{equation}
   \sum_{\kappa=1}^{l}p(\kappa|\bmlambda)(\Tilde{\lambda}_{i}-\lambda_{i}) = 0, \quad \sum_{\kappa=1}^{l} \Tilde{\lambda}_{i}\, {\partial_j p(\kappa|\bmlambda)} = \delta_{ij},
\end{equation}
where $\partial_j = {\partial}/{\partial \lambda_{j}}$. Which provides the classical Cramér-Rao bound, puts a constraint on all possible mean-square error matrices, and can be written as 
\begin{equation}
    \mathcal{O}(\bmlambda) \geq \frac{\F^{-1}}{N}, \label{classicalCRB}
\end{equation}
where $\F$ is the classical Fisher information matrix with elements
\begin{equation}
    F_{ij} = \sum_{\kappa}p(\kappa|\bmlambda)\big(\partial_{i} \log p(\kappa|\bmlambda)\big)\big(\partial_{j} \log p(\kappa|\bmlambda)\big).
\end{equation}
This bound is valid for any fixed classical statistical model $p(\kappa|\bmlambda)$ and can always be achieved using locally unbiased estimators for any $N$. For more practical estimators, such as the maximum likelihood estimator \cite{hradil04,d2000parameter}, the Cramér-Rao bound is asymptotically reached as the number of measurements $N$ approaches infinity ($N\rightarrow\infty$). In this scenario, the mean square error matrix becomes proportional to the inverse of the Fisher information matrix, scaled by ${1}/{N}$.

In order to calculate the quantum Fisher information, one introduces the symmetric logarithmic derivative (SLD) operator $\L_{i}^S$ for each parameter \cite{1976Helstrom,Yuen}, which is defined by the expression as 
\begin{equation}
\partial_{i}\hat{\rho}_{\bmlambda} = \frac{\L_{i}^S\hat{\rho}_{\bmlambda}+\hat{\rho}_{\bmlambda}\L_{i}^S}{2}. \label{SLD}
\end{equation}
From Eq. \eqref{SLD}, Fisher information matrix elements can be written as
\begin{equation}
    Q_{ij} = \hbox{Tr}\Bigg[\hat{\rho}_{\bmlambda}\frac{\L_{i}^S\L_{j}^S+\L_{j}^S \L_{i}^S}{2}\Bigg],
\end{equation}
which provides the measurement-independent quantum CRB matrix
\begin{equation}
    \mathcal{O}(\bmlambda) \geq \Q^{-1} \label{QCRB}
\end{equation}
where $\Q$ is the so-called quantum Fisher information matrix and we drop the factor $N$ from the above and forthcoming equations.

Inequality \eqref{QCRB} is, in general, not saturable for multi-parameter models. This is a simpler case of one-parameter quantum statistical models, for which the single SLD operator defines the best measurement, i.e. the one providing the highest FI value, and the bound can in turn always be saturated. Other QFI matrices can be defined by using other definitions of the logarithmic derivative, but none of them provides a tight bound that can be saturated for the multi-parameter case. Its general non-attainability can be understood by considering parameters whose SLD operators, corresponding to optimal measurement methods, do not commute. To gain deeper insights into the performance of different multi-parameter estimators, it is common practice to convert matrix bounds into scalar bounds.

All those matrix inequalities give rise to scalar bounds by introducing a $l \times l$, real and positive weight matrix $\W$. On taking the trace of Eqs. \eqref{classicalCRB} and \eqref{QCRB} with weight matrix $\W$, we can write
 \begin{equation}
    \Tr[\W\mathcal{O}]\geq C_{F},~~\Tr[\W\mathcal{O}]\geq C_{Q},
 \end{equation}
where,
\begin{equation}
    \begin{split}
        C_{F} = \Tr\big[\W\F^{-1}\big],~~ C_{Q} = \Tr\big[\W \Q^{-1}\big]\label{C_{Q}},
    \end{split}
\end{equation}
are the scalar classical and quantum CRB, respectively. Similarly, the corresponding matrix bound, these scalar bounds are in general not attainable. Therefore, we can use the technique of considering minima of the classical scalar bound over all the possible POVMs and termed it as the {\em most informative bound},
\begin{equation}
    C_{M\!I} = \min_{\hat{\Pi}} \Tr\big[\W \F^{-1}\big]\,.
\end{equation}
The quantity $C_{M\!I}$ does not, in general, coincide with $C_{Q}$ in the multi-parameter case. If we perform the measurement on infinitely many copies of the statistical model, we can achieve the so-called { \em Holevo Cramér-Rao bound} (HCRB) $C_{H}$. In summary, we have,
\begin{equation}
    \Tr[\W\mathcal{O}] \geq C_F \geq C_{M\!I} \geq C_H \geq C_Q.
\end{equation}
Therefore, the tightest bound, at least asymptotically and with collective strategies on multiple copies of the unknown state, is provided by the Holevo bound.
 The Holevo bound $C_H$ is usually difficult to evaluate compared to $C_Q$ and therefore the following relation represents a useful tool in characterizing the a multi-parameter estimation model
\begin{equation}
C_{Q} \leq C_H \leq (1+\mathcal{R}) C_Q \label{C_H(W)},
\end{equation}
where the {\em quantumness} parameter $\mathcal{R}$ is given by 
 \begin{equation}
     \mathcal{R} = ||i\, \Q^{-1} \, \U||_{\infty}\,,
 \end{equation}
where, $||\X||_{\infty}$ denotes the largest eigenvalue of the matrix $\X$ and $\U$ denotes the asymptotic incompatibility matrix, commonly known as the {\em Uhlmann curvature}, having matrix elements 
 \begin{equation}
     U_{\mu\nu} = -\frac{i}{2}\Tr\big[\hat{\rho}_{\bmlambda}(\L_{\mu}\L_{\nu} - \L_{\nu}\L_{\mu})\big].
 \end{equation}
The quantumness parameter $\mathcal{R}$ is bounded $0 \leq \mathcal{R} \leq 1$ and vanishes $\mathcal{R} = 0$ iff  $\U$ is the null matrix and the QFI bound may be saturated, which is usually referred to as the {\em weak compatibility condition}.  Therefore, it provides a measure of asymptotic incompatibility between the parameters. 
\textcolor{black}{
A relevant property of the quantumness parameter $\mathcal{R}$ is that it does not depend on the weight matrix $\W$, such that the bound in Eq. (\ref{C_H(W)}) is valid for any $\W$. 
On the other hand, it is often the case that a natural weight matrix emerges from the physical properties of the quantum statistical model. In this case, a weight-dependent 
chain of inequalities for the Holevo bound $C_H$ may be obtained, which is tighter 
than that in Eq.(\ref{C_H(W)}). We have
\begin{equation}
C_{Q} \leq C_H \leq (1+\mathcal{T}) C_Q \label{Tbrac}\,,
\end{equation}
where 
\begin{equation}
\mathcal{T}_W = \frac{||\sqrt{\W} \Q^{-1}\U \Q^{-1}\sqrt{\W}||_1}{C_{Q}}
\label{Tdel}\,,
\end{equation}
and 
$||\X||_{1}$ denotes the sum of the singular values of $\X$ \cite{PhysRevLett.123.200503}.
In our case, the two hyperbolic phase should be treated on an equal foot, and the natural
weight matrix is the identity. By this choice and exploiting the fact that we have 2 parameters
we may write
 \begin{align}
     \mathcal{R} & = \sqrt{\frac{\mathcal{S}}{\mathcal{C}}}, \label{R} \\[1ex]
       \mathcal{T_{\mathbb{I}}} & = \sqrt{\frac{2}{\mathcal{C}}} \frac{1}{\Tr[\Q]}, \label{T}
 \end{align}
where,
\begin{equation}\label{scdef}
    \mathcal{C} = \frac{1}{\det[\U]}~~\text{and}~~\mathcal{S} = \frac{1}{\det[\Q]},
\end{equation}
are the incompatibility and sloppiness quantifiers, respectively. 
Since $\Q$ is a positive definite matrix, we have $2/\hbox{Tr}[\Q] \leq 1/\sqrt{\hbox{det}[\Q]}$ and therefore $\mathcal{T_{\mathbb{I}}} \leq  \mathcal{R}$.}
\par
\textcolor{black}{The explicit evaluation of the above bounds may be carried out analytically for pure  models, where the QFI matrix elements can be expressed as}
\begin{equation}
    \begin{split}
Q_{jk} = 4 \,\text{Re}\big[\big\langle \partial _j\psi \big|\partial _k\psi \big\rangle - \big\langle \partial _j\psi \big|\psi \big\rangle \big\langle \psi \big|\partial _k\psi \big\rangle \big], \label{Q}
\end{split}
\end{equation}
and the analogous expression for the elements of the Uhlmann curvature is given by
\begin{equation}
    \begin{split}
U_{jk} = 4 \,\text{Im}\big[\big\langle \partial _j\psi \big|\partial _k\psi \big\rangle -\big\langle \partial _j\psi \big|\psi \big \rangle \big \langle \psi \big|\partial _k\psi \big\rangle \big]. \label{U}
\end{split}
\end{equation}

\section{Scrambling information in two-parameter hyperbolic phase estimation}
\label{Proposed model}
\begin{figure}[h!]
\centering
\includegraphics[width=0.7\columnwidth]{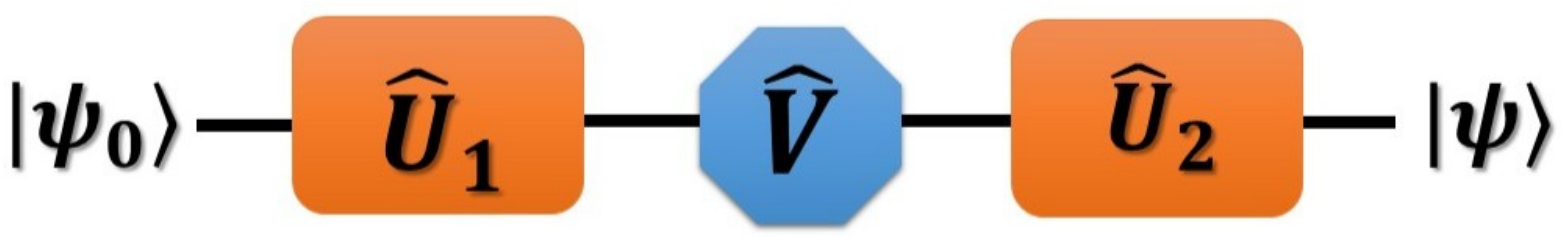}
\caption{\label{MPA} Schematic diagram of our proposed model to reduce the sloppiness in the 
estimation of two successive squeezings. Here, $\hU_1$ and $\hU_2$ are the unitary operators encoding the  parameters $\lambda_1$ and $\lambda_2$, respectively. $\hV$ is the scrambler with the phase $\phi$ to decouple the parameters.}
\end{figure}

Let us consider the scheme Fig. \ref{MPA}: we consider two unitary operators defined as
\begin{equation}
 \hU_{k} = \text{exp}\Bigg({- \frac{i}{2}\,\lambda_k ~\hG}\Bigg)\,, \quad k=1,2\,,\label{unitaryop}
\end{equation}
where $\lambda_{k}$ denote the encoded parameters to be estimated 
and $\hG = \hat{a}^2 + \hat{a}^{\dagger 2}$ is 
the squeezing generator with $\hat{a}$ ($\hat{a}^{\dagger}$) as the photon annihilation (creation) operator. If no operation is performed between the two unitaries, then the output state will depend only on the sum of the two parameters $\lambda_1$ and $\lambda_2$, which cannot be estimated separately. The model is thus {\em sloppy}.
 
To decouple the parameters, we introduce a {\em scrambling} operation in between the unitary operations $\hU_1$ and $\hU_2$
 \begin{equation}
 \hV = \text{exp}\big(-i\phi\, \hat{a}^{\dagger}\hat{a}\big),
 \end{equation}
where $\phi$ is a tunable parameter.  The scrambler makes it possible to encode the 
two parameters independently, making it possible to jointly estimate both of them. 

For the probe initially prepared in a coherent state $|\alpha\rangle $, 
where $|\alpha| \in \mathbb{R}^+$ is the amplitude and $\theta$ is the phase, the
final state can be written as
\begin{equation}
|\psi \rangle =\hU_2 \hV \hU_1|\alpha\rangle.\label{psi}
\end{equation}
Our task is now to calculate the QFI and Uhlmann matrix elements defined 
in Eqs. \eqref{Q} and \eqref{U}. We can write
\begin{equation}
    |\partial_1\psi\rangle = -\frac{i}{2}\hU_2\hV\hU_1G |\alpha \rangle,
\end{equation}
\begin{equation}
|\partial_2\psi\rangle = -\frac{i}{2}\hG\hU_2\hV\hU_1 |\alpha \rangle\,,
\end{equation}
where we used the unitary operator defined as in Eq. \eqref{unitaryop}. The QFI matrix elements 
read as follows
\begin{subequations}
\begin{align}
Q_{11} =&\ 4 \left[\big\langle\alpha\big|\hG^2\big|\alpha \big\rangle -\big\langle\alpha\big|\hG\big|\alpha \big\rangle^2\right] \\
Q_{12} =&\, 4\, \hbox{Re}\left[\big\langle\alpha\big|\hG\hU_1^{\dagger}\hV^{\dagger}\hG\hV\hU_1
\big|\alpha \big\rangle \right.\notag \\ & 
\left.  \hspace{0.5cm}- \big\langle\alpha\big|\hG\big|\alpha \big\rangle 
\big\langle\alpha\big|\hU_1^{\dagger}\hV^{\dagger}\hG\hV\hU_1\big|\alpha \big\rangle \right]\label{QFIelements}\\
Q_{21} =&\, Q_{12} \\ 
Q_{22} =&\, 4 \, \left[\big\langle\alpha\big|\hU_1^{\dagger}\hV^{\dagger}\hG^2\hV\hU_1\big|\alpha \big\rangle 
\right. \notag\\
& \left. \hspace{0.5cm}-\big\langle\alpha\big|\hU_1^{\dagger}\hV^{\dagger}\hG\hV\hU_1\big|\alpha \big\rangle^2\right]\,, 
\end{align}
\end{subequations}
where $\hU_j^{\dagger}\hU_j = \hU_j\hU_j^{\dagger} = \mathbbm{1}$ and $\hV^{\dagger}\hV = \mathbbm{1}$. The elements of the Uhlmann curvature matrix become
\begin{subequations}\label{Ulmanelements}
\begin{align}
U_{11} =&\ U_{22} = 0 \\
U_{21} =& -U_{12} = 4 \, \text{Im}\big[\big\langle\alpha\big|\hG\hU_1^{\dagger}\hV^{\dagger}\hG\hV\hU_1\big|\alpha \big\rangle  \notag\\
& \hspace{2cm} -\big\langle\alpha\big|\hG\big|\alpha \big\rangle 
\big\langle\alpha\big|\hU_1^{\dagger}\hV^{\dagger}\hG\hV\hU_1\big|\alpha \big\rangle \big] . 
\end{align}
\end{subequations}
Using Eqs. \eqref{QFIelements} and \eqref{Ulmanelements}, we can calculate the 
incompatibility and sloppiness parameters, and using Eq. \eqref{R} and Eq. \eqref{T} we 
evaluate $\mathcal{R}$ and $\mathcal{T_{\mathbb{I}}}$, respectively.

For calculating the lower bound, we need to calculate $C_{Q}$ given in Eq. \eqref{C_{Q}}. For $\W = \mathbbm{1}$, $C_{Q}$ can be written as
\begin{equation}
   C_{Q} = \frac{Q_{11}+Q_{22}}{Q_{11}Q_{22}-Q_{12}Q_{21}} = \frac{\Tr[\Q]}{\det[\Q]}. \label{Cq}
\end{equation}
Our task is to reduce the dependency on parameters to measure both parameters simultaneously and precisely. As we can see from Eqs. \eqref{QFIelements} and \eqref{Ulmanelements}, the matrix elements of the QFIM and of the Uhlmann curvature, are independent of $\lambda_2$. This means that
$C_Q$, $\mathcal{S}$ and $\mathcal{R}$ are independent on $\lambda_2$.
\begin{figure}[h!]
\centering
\includegraphics[width=.7\columnwidth]{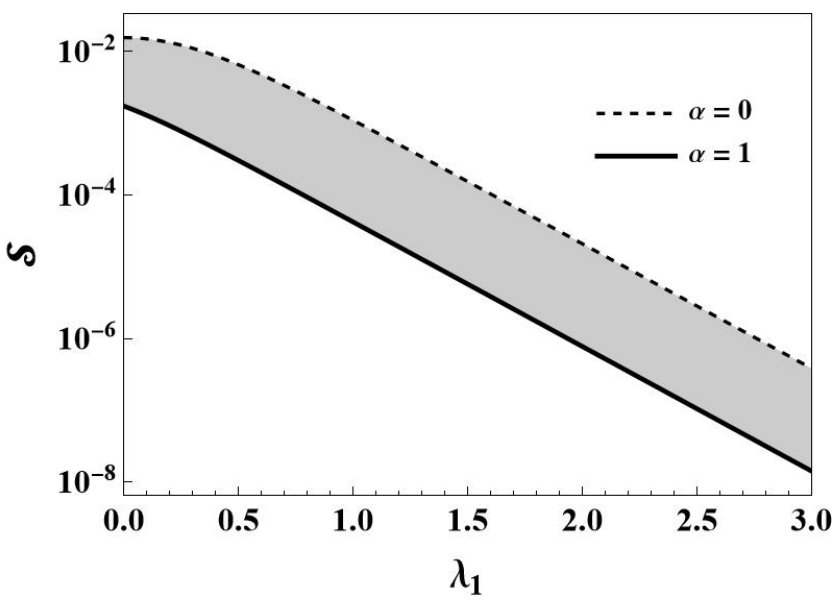}
\caption{\label{Sl} The sloppiness $\mathcal{S}$ as a function of $\lambda_1$ for $\alpha = 0, 1$. 
The phases are set $\phi = \pi/4, ~\theta = \pi/4$. Dashed line denotes results for $\alpha = 0$ and solid line for $\alpha = 1$, respectively.}
\end{figure}

Upon inspecting analytically the expression of sloppiness, we found that  
$\phi = \theta = \pi/4$ provide the minimum value of $\mathcal{S}$ at fixed values of the other quantities. This is given by 
\begin{align}
\mathcal{S} = & \frac{1}{8} \Bigg[ \frac{1}{1 + 2 \alpha^2}  + \frac{2}{1 + (1 + 4 \alpha^2) \cosh(4 \lambda_1) + 4 \alpha^2 \sinh(4 \lambda_1)} \Bigg]\,,
 \end{align}
where, without loss of generality, we have assumed $\alpha \in {\mathbbm R}$ and $\alpha \ge 0$.

Those values will be 
employed from now on in further analysis of the model. In Fig. \ref{Sl}, we illustrate
the behaviour of the sloppiness $\mathcal{S}$ as a function of $\lambda_1$ for $\alpha = 0$ and $\alpha=1$ (the black region corresponds to intermediate values). We can see that as we 
increase the value of 
$\lambda_1$ the sloppiness decreases, and this is true also by increasing the 
coherent amplitude.

\begin{figure}[h!]
\centering
\includegraphics[width=.7\columnwidth]{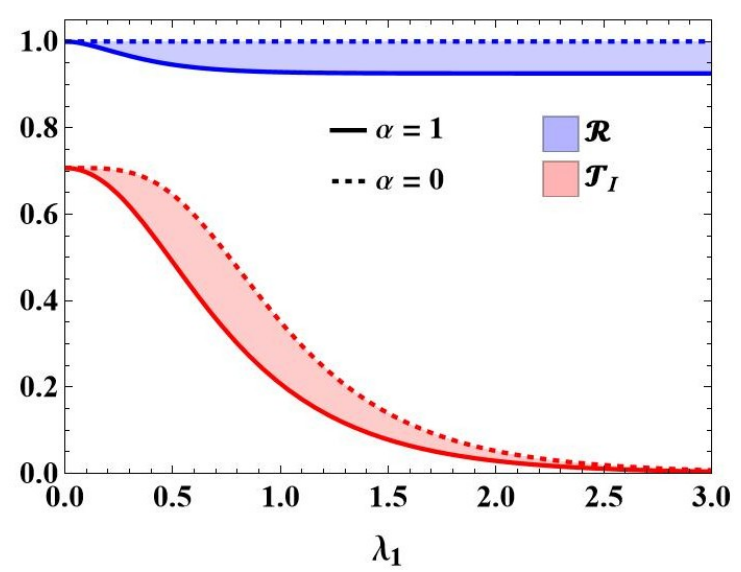}
\caption{\label{Rl} The quantities $\mathcal{R}$ and $\mathcal{T_\mathbb{I}}$ as functions
of $\lambda_1$ for t$\alpha = 0, 1$. Blue region shows the $\mathcal{R}$ values while red one shows $\mathcal{T_\mathbb{I}}$. The dashed lines denote results for $\alpha = 0$ and solid line for $\alpha = 1$, respectively. The phases are set to $\phi = \pi/4, ~\theta = \pi/4$.}
\end{figure}

Eq. \eqref{C_H(W)} says that the Holevo bound lies between $C_Q$ and $C_Q(1+\mathcal{R})$. 
In Fig. \ref{Rl}, to illustrate how the gap between $C_Q$ and $C_Q(1+\mathcal{R})$ 
depends on the relevant parameters, we show the behaviour of $\mathcal{R}$ as a function 
of $\lambda_1$ for $\alpha = 0$ and $\alpha =1$. For a vacuum probe, $\mathcal{R}$ is 
independent on $\lambda_1$, while for nonzero coherent amplitude it does depend 
on $\lambda_1$, but quickly saturates for increasing $\lambda_1$. For large $\alpha$ we have
\begin{align}
\mathcal{R} & = 1 - \frac{2}{\alpha^2} e^{-4 \lambda_1} \sinh^2 \lambda_1\,\cosh^2 \lambda_1+ O(1/\alpha^3) \notag \\
& \simeq 1 - \frac{1}{8 \alpha^2}+ O(1/\alpha^3) 
\end{align}

The quantifier ${\cal R}$ provides a general, weight-matrix independent, bound to the Holevo limit. 
However, when the parameters are well identified and their relevance is known, a more tight 
weight-matrix dependent bound may be introduce. This is given in terms of the parameter $\mathcal{T}_W$ defined in Eq. \eqref{Tdel}. In this case, Eq. \eqref{Tbrac} provides the limits for Holevo bound. Since our parameters are equally important, we consider $\W = \mathbb{I}$ and study the dependence of 
$\mathcal{T}_{\mathbb{I}}$ on $\alpha$ and $\lambda_1$. In Fig \ref{Rl}, we show the behaviour 
of $\mathcal{T}_{\mathbb{I}}$ as a function of $\lambda_1$ for $\alpha = 0$ and $\alpha =1$. In both cases, $\mathcal{T_\mathbb{I}}$ decreases quite quickly with $\lambda_1$. For large $\alpha$, we have 
$$\mathcal{T_\mathbb{I}} \simeq \frac{1}{\sqrt{2}\, \cosh (2 \lambda_1)}
\,.$$ 

\begin{figure}[h!]
\centering
\includegraphics[width=.7\columnwidth]{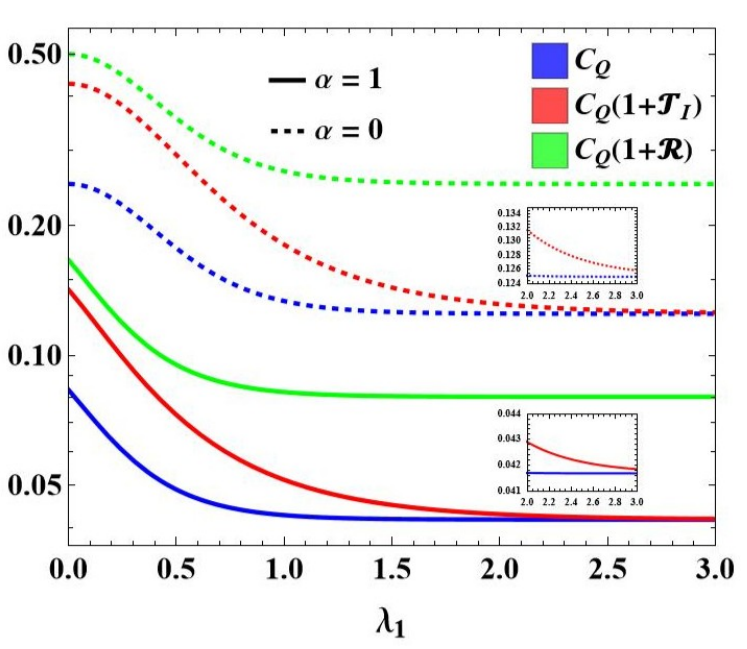}
\caption{\label{Cq_L} The bounds $C_{Q}$, $C_{Q}(1+\mathcal{T_\mathbb{I}})$ and $C_{Q}(1+\mathcal{R})$  as a function of $\lambda_1$ for  $\alpha = 0,1,~\phi = \pi/4,~\theta = \pi/4$. Dashed lines denote
the bounds for $\alpha = 0$ and solid lines are for $\alpha = 1$. The insets show results for large $\lambda_1.$}
\end{figure}

Let us now illustrate the behaviour of the various bounds, i.e., 
$C_Q$, $C_{Q}(1+\mathcal{T_\mathbb{I}})$ and  $C_Q(1+\mathcal{R})$ as a function of 
$\lambda_1$. This is shown in Fig. \ref{Cq_L} for $\alpha = 0$  and $\alpha =1$ . As it is apparent from the plot, increasing $\lambda_1$ the regions between $C_Q$ and $C_Q(1+\mathcal{R})$ become 
narrower, and this phenomenon is even more accentuated for the regions between 
$C_Q$ and
$C_{Q}(1+\mathcal{T_\mathbb{I}})$, which become negligible for large $\lambda_1$ 
(as shown in the insets).

We conclude that when the true value of $\lambda_1$ is large enough, the ultimate bound to 
precision joint estimation of the two hyperbolic phase-shift coincides with the SLD bound 
$C_{Q}$, which itself is achievable by measurements performed on repeated preparations of 
the probe, independently on its amplitude. On the other hand, for lower values of 
$\lambda_1$ (a regime which may be of interest in certain situations \cite{PARIS1995132}), the gap  between $C_{Q}$, and $C_{Q}(1+\mathcal{T_\mathbb{I}})$ is larger and 
reaching the ultimate precision bound likely requires collective (entangled) measurements on a 
collection of repeated preparations.

\section{Stepwise estimation of parameters}\label{separate measurement}
\begin{figure}[h!]
\centering
\includegraphics[width=0.7\columnwidth]{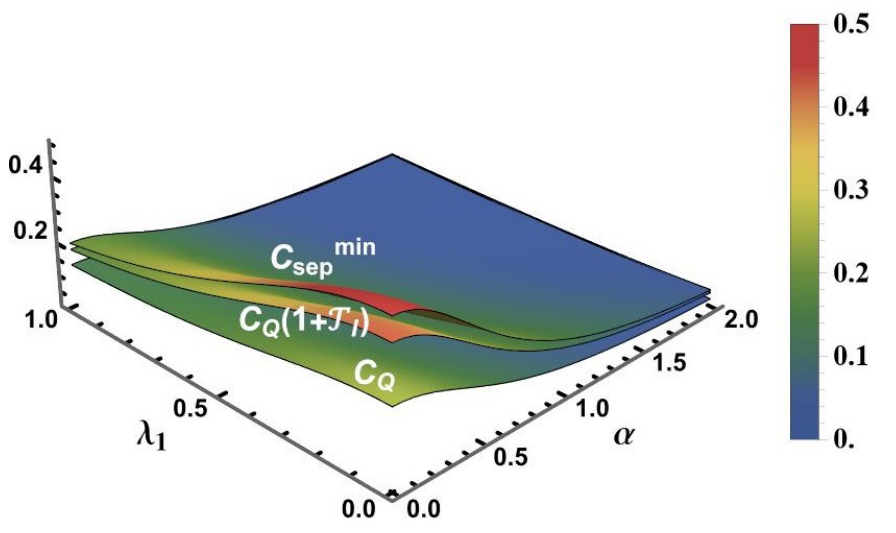}
\caption{\label{Csep12} The optimized stepwise 
bounds $C_{\sep_1}$ and $C_{\sep_2}$ compared to $C_{Q}$, $C_{Q}(1+\mathcal{T_{\mathbb I}})$ 
as a function the parameter $\lambda_1$ and the coherent probe amplitude $\alpha$.}
\end{figure}

\textcolor{black}{In the previous Sections, we have addressed the joint estimation of the two parameters 
$\lambda_1$ and $\lambda_2$. Now we are interested in comparing the precision of those strategies
with that resulting from {\em stepwise} strategies, i.e., protocols where the available 
copies of the probe state are split in two sets and then used to estimate only one of the two parameters \cite{stepwise}. 
In conducting separate measurements, we can opt for two different approaches: 
measuring $\lambda_1$ first and then $\lambda_2$, or we can start with 
$\lambda_2$ and then estimate $\lambda_1$.
Let us denote by $M$ the total number of available copies, of which $\gamma M$ copies are
exploited to estimate $\lambda_1$ and $(1-\gamma)M$ for $\lambda_2$, with $0\leq\gamma\leq1$. 
If we estimate $\lambda_1$ first, the precision bound can be obtained from the 
Eq. \eqref{C_{Q}} and for a diagonal weight matrix $\W=\hbox{Diag}(1,0)$, i.e., 
\begin{equation}
    \Delta\lambda_1^2 \geq \frac{[Q^{-1}]_{11}}{\gamma M}.
\end{equation}
where $[Q^{-1}]_{11}$ denotes the first diagonal element of the inverse QFIM, $\Q^{-1}$. If we now 
proceed to estimate $\lambda_2$, we can take into account the information gained 
about $\lambda_1$ in the first step. As a consequence, this second step becomes a 
single parameter estimation task, and the precision bound can be written as
\begin{equation}
    \Delta\lambda_2^2 \geq \frac{1}{Q_{22}(1 - \gamma) M}.
\end{equation}
The total variance in this strategy would be the sum of both measurements, i.e., $\Delta\lambda_1^2 + \Delta\lambda_2^2$, which is bounded by $C_{\sep_1}/M$ where 
\begin{equation}
C_{\sep_1} = \frac{[Q^{-1}]_{11}}{\gamma} +\frac{1}{Q_{22}(1-\gamma)}\,.
\end{equation}
Similarly in the second scenario, we first estimate $\lambda_2$ and then $\lambda_1$, 
and the total variance is bounded by $C_{\sep_1}/M$, where 
\begin{equation}
C_{\sep_2} = \frac{[Q^{-1}]_{22}}{\gamma} +\frac{1}{Q_{11}(1-\gamma)}\,,
\end{equation}
where $\gamma$ in both formulas is the fraction of measurements used in the first step.
Upon inserting the explicit expression of the elements of the inverse QFIM (as functions of the $\Q$ matrix elements) we have
\begin{align}
C_{\sep_1} & = \frac{{\cal S}\, Q_{22}}{\gamma} +\frac{1}{Q_{22}(1-\gamma)}\,, \label{csep1} \\[1ex]
C_{\sep_2} & = \frac{{\cal S}\, Q_{11}}{\gamma} +\frac{1}{Q_{11} (1-\gamma)}\,,\label{csep2}
\end{align}
where ${\cal S}$ is the sloppiness of the model, defined in Eq. (\ref{scdef})  and the values 
of $Q_{11}$ and $Q_{22}$ are given in Eq. \eqref{QFIelements}. 
The value of $\gamma$  may be tuned to minimize the quantity $C_{\sep_k}$, this is obtained 
choosing $$\gamma_k = \frac{Q_{kk} \sqrt{{\cal S}}}{1 + Q_{kk} \sqrt{{\cal S}}}\,,$$ 
leading to
\begin{align}
C_{\sep_k}^{\rm min} & = \frac{1}{Q_{kk}} + Q_{kk} {\cal S} +  2 \sqrt{{\cal S}}\,. \label{csepmin} 
\end{align}
To compare the performance of the separate strategy to that of the joint one,  in Fig. \ref{Csep12} we show the optimized value of $C_{\sep_1}$ and $C_{\sep_2}$ (they are equal) alongside $C_{Q}$ and $C_{Q}(1+\mathcal{T_{\mathbb I}})$, as a function the parameter $\lambda_1$ and the coherent probe amplitude $\alpha$ (there is no dependence on the value of the second parameter $\lambda_2$). 
We observe that the stepwise bounds, even when optimized, are larger than $C_{Q}(1+\mathcal{T_{\mathbb I}})$ in the entire region. Hence, we conclude that joint estimation is always better than separate one for two-parameter estimation in our model.}

\section{Phase-space analysis}\label{s:phase:space}
A further insight into our results can be obtained considering the phase-space description of the evolved state after the two squeezing operations \cite{olirev2012}.
If we introduce the quadrature operators:
\begin{align}\label{fm:def}
\hat q= \frac{\hat a+\hat a^\dag}{\sqrt{2}},\quad
\hat p= \frac{\hat a-\hat a^\dag}{i\sqrt{2}},
\end{align}
we can define the first moment vector:
\begin{equation}
    \boldsymbol{X} =
\big(\langle \hat q\rangle,
\langle \hat p\rangle \big)^\intercal
\end{equation}

the covariance matrix:
\begin{equation}\label{CM:def}
    \boldsymbol{\sigma}=\left(
\begin{array}{cc}
\Delta q^2 & \Delta qp \\[1ex]
\Delta qp & \Delta p^2
\end{array}
    \right)
\end{equation}
where $\Delta q^2 = \langle \hat q^2\rangle - \langle \hat q\rangle^2$ and a similar definition holds for $\Delta p^2$, while $\Delta qp = \langle \hat q\hat p + \hat p\hat q\rangle/2 - \langle \hat q\rangle\langle \hat p\rangle$.

It is worth noting that the covariance matrix of an optical state can be experimentally measured resorting, for instance, to the homodyne detection \cite{olirev2021}.

In the presence of the output state (\ref{psi}) with $|\psi_0\rangle = |\alpha e^{i\theta} \rangle$, $\alpha \in {\mathbbm R}$, we have:
\begin{subequations}
\begin{align}
\langle \hat q\rangle &= \sqrt{2}\, \alpha\, e^{\lambda_2}\Big( e^{\lambda_1} \cos\theta\, \cos\phi - e^{-\lambda_1} \sin\theta\, \sin\phi \Big),\\[1ex]
\langle \hat p\rangle &= -\sqrt{2}\, \alpha\, e^{-\lambda_2}\Big( e^{-\lambda_1} \sin\theta\, \cos\phi + e^{\lambda_1} \cos\theta\, \sin\phi \Big),
\end{align}
\end{subequations}
and
\begin{subequations}
\begin{align}
\Delta q^2 &= \frac{e^{2 \lambda_2}}{2}\Big(
e^{2\lambda_1} \cos^2\phi + e^{-2\lambda_1} \sin^2\phi
\Big),\\[1ex]
\Delta p^2 &= \frac{e^{-2 \lambda_2}}{2}\Big(
e^{-2\lambda_1} \cos^2\phi + e^{2\lambda_1} \sin^2\phi
\Big),\\[1ex]
\Delta qp &= - \frac12 \, \sin (2\phi)\, \sinh (2\lambda_1)\,.
\end{align}
\end{subequations}
We easily see that if we set $\phi=0$ the first moment vector and the covariance matrix elements reduce to:
\begin{subequations}
\begin{align}
\langle \hat q\rangle &= \sqrt{2}\, \alpha\, e^{\lambda_1+\lambda_2} \cos\theta,\\[1ex]
\langle \hat p\rangle &= - \sqrt{2}\, \alpha\, e^{-(\lambda_1+\lambda_2)} \sin\theta,
\end{align}
\end{subequations}
and:
\begin{align}
\Delta q^2 = \frac{e^{2 (\lambda_1 + \lambda_2)}}{2},\,\,
\Delta p^2 = \frac{e^{-2 (\lambda_1 + \lambda_2)}}{2},\,\,
\Delta qp = 0,
\end{align}
respectively,
while for $\phi=\pi/2$ we find:
\begin{subequations}
\begin{align}
\langle \hat q\rangle &= -\sqrt{2}\, \alpha\, e^{\lambda_2-\lambda_1} \sin\theta,\\[1ex]
\langle \hat p\rangle &= - \sqrt{2}\, \alpha\, e^{-(\lambda_2-\lambda_1)} \cos\theta,
\end{align}
\end{subequations}
and:
\begin{align}
\Delta q^2 = \frac{e^{2 (\lambda_2 - \lambda_1)}}{2},\,\,
\Delta p^2 = \frac{e^{-2 (\lambda_2 - \lambda_1)}}{2},\,\,
\Delta qp = 0.
\end{align}
In both these cases, the cross-correlation term $\Delta qp$ vanishes and the expectations of the quadrature operators as well as their variances $\Delta q^2$ and $\Delta p^2$ come to depend only on the sum, $\lambda_1+\lambda_2$, or difference, $\lambda_1-\lambda_2$, of the two unknown parameters, thus allowing their joint estimation. Moreover, now the absolute value of the cross-correlation term reaches its maximum.

\section{Joint estimation of the hyperbolic phases through general-dyne measurement}\label{generaldyne}

A general-dyne measurement is described by the POVM:
\begin{equation}
\hat\Pi_z(\beta) = \frac{\hat D^\dag(\beta)\hat \rho_{\rm m}(z)\hat D(\beta)}{\pi},
\end{equation}
where $\beta\in {\mathbbm C}$ is related to the measurement outcome and $\hat \rho_{\rm m}(z)$ is a pure Gaussian state with null first moment vector and covariance matrix, defined as in Eq.~(\ref{CM:def}), given by:
\begin{equation}
    \boldsymbol{\sigma}_{\rm m}= \frac12 \left(
\begin{array}{cc}
z & 0 \\[1ex]
0 & z^{-1}
\end{array}
    \right),
\end{equation}
with $z\in {\mathbbm R}$, $z\ge 0$. The general-dyne measurement reduces to the well-known double homodyne (heterodyne) one if $z=1$, whereas in the limits $z\to 0$ and $z\to \infty$ it leads to the homodyne detection POVM associated with the measurement of the quadratures $\hat q$ and $\hat p$, respectively. In our scenario, both the measured state and the measurement are Gaussian, thus outcome statistics is Gaussian, too. As usual, the collected data sample can be processed to perform the joint estimation of the two involved hyperbolic phases, $\lambda_1$ and $\lambda_2$ through suitable unbiased estimators. In the following, we use the Fisher information matrix as a figure of merit to assess the performance of the particular measurement we have chosen.

The Fisher information matrix $\boldsymbol{F}$ can be written as a function of the state first moment vector $\boldsymbol{X}$ and of the  state covariance matrix $\boldsymbol{\sigma}$ introduced in Sect.~\ref{s:phase:space}. The  Fisher information matrix elements are given by:
\begin{align}
F_{jk} &=
\left( \frac{\partial\boldsymbol{X}^\intercal}{\partial \lambda_j} \right)
\boldsymbol{\Sigma}^{-1}
\left( \frac{\partial\boldsymbol{X}}{\partial \lambda_k} \right) + \frac12\, \mbox{Tr}\left[
\boldsymbol{\Sigma}^{-1}
\left( \frac{\partial\boldsymbol{\Sigma}}{\partial \lambda_j} \right)
\boldsymbol{\Sigma}^{-1}
\left( \frac{\partial\boldsymbol{\Sigma}}{\partial \lambda_k} \right)
\right]
\end{align}
where $\boldsymbol{\Sigma} = \boldsymbol{\sigma} + \boldsymbol{\sigma}_{\rm m}$ is the covariance matrix of the measured statistics.

Starting from $\boldsymbol{F}$, we can obtain the following classical bound:
\begin{equation}
C_{\rm g} = \mbox{Tr}\big[\boldsymbol{F}^{-1}\big],
\end{equation}
which depends on $\lambda_1$, $\lambda_2$, but also on the probe parameters $\alpha$ and $\theta$, on the scrambling phase $\phi$ and on the general-dyne parameter $z$. The analytical expression of $C_{\rm g}$ is rather cumbersome and it is not reported here explicitly. However, it can be optimized, namely, minimized, with respect to $\theta$, $\phi$ and also to $z$. The optimization leads to the choice $\theta = 0$, $\phi= \pi/4$ and $z=\exp(2\lambda_2)$. Indeed, the last condition on the general-dyne parameters requires resorting to an adaptive method to first retrieve the value of the hyperbolic phase $\lambda_2$. Eventually, the optimized Fisher information, $\boldsymbol{F}^{\rm (opt)}$ is independent of $\lambda_2$, as in the case of the QFI matrix investigated in Sect.~\ref{Proposed model}, it is diagonal, i.e., $F^{\rm (opt)}_{12} = F^{\rm (opt)}_{21} = 0$, and the non-null elements reads:
\begin{subequations}
\begin{align}
F^{\rm (opt)}_{11} &= 1 + \tanh^2 \lambda_1  +2\alpha^2 \left[1+\tanh \lambda_1 \right],\\[1ex]
F^{\rm (opt)}_{22} &= \frac{2\alpha^2 \,e^{3\lambda_1}}{\cosh \lambda_1 } + \frac{\cosh^2 (2\lambda_1) }{\cosh\lambda_1},
\end{align}
\end{subequations}
thereafter one can calculate the bound $C_{\rm g}$.

As one may expect, the comparison with the quantum bounds retrieved in Sect.~\ref{Proposed model} leads to $C_{\rm g} > C_Q$ , $\forall \lambda_1, \alpha$, but it is interesting to investigate the limit of high excited coherent probe, i.e., $\alpha \gg 1$, where we find:
\begin{equation}
C_{\rm g} \simeq \frac{\left(1+e^{-4\lambda_1}\right)^2}{4\alpha^2}\qquad (\alpha \gg 1)
\end{equation}
which can be compared to the the lower and upper bound of $C_{H}$, namely:
\begin{align}
&C_{Q}  \simeq \frac{1+e^{-4\lambda_1}}{16\alpha^2}\qquad (\alpha \gg 1)\\[1ex]
&C_{Q}(1+{\cal R}) \simeq \frac32 \,C_{Q} \qquad (\alpha \gg 1)
\end{align}
respectively. After some calculation we obtain (still in the limit $\alpha \gg 1$):
\begin{equation}
\frac{1}{4\left(1+e^{-4\lambda_1}\right)} \lesssim
\frac{C_H}{C_{\rm g}} \lesssim
\frac{3}{8\left(1+e^{-4\lambda_1}\right)}
\end{equation}
which, for $\lambda_1 \gg 1$, reduces to:
\begin{equation}
0.25 \lesssim \frac{C_H}{C_{\rm g}} \lesssim 0.37
\quad \mbox{if} \quad \alpha, \lambda_1 \gg 1.
\end{equation}

\textcolor{black}{If we focus on the case where the weight matrix is the identity we then
have
\begin{align}
T_{\cal I} \simeq \frac{1}{\sqrt{1+ \cosh (4 \lambda_1)}}\qquad (\alpha \gg 1)\,,
\end{align}
meaning that $C_H\simeq C_Q$ for large $\lambda_1$ and thus
\begin{equation}
\frac{C_H}{C_{\rm g}} \simeq \frac14
\quad \mbox{if} \quad \alpha, \lambda_1 \gg 1.
\end{equation}
}

\section{Conclusion }\label{Conclution}
In conclusion, we have analyzed in detail an interaction scheme involving two successive unknown squeezings (hyperbolic phase shifts), which are the quantities to be estimated. The quantum statistical model is inherently sloppy, and the quantum Fisher information matrix becomes singular. However, estimating both parameters becomes feasible if an operation is performed between the two squeezings to scramble information across the Hilbert space.

In particular, we have considered the effects of a phase-shift scrambling transformation, optimized to reduce the model’s sloppiness and maximize overall estimation precision. We found that when the true value of the first squeezing, $\lambda_1$, is sufficiently large, the ultimate bound on joint estimation precision coincides with the SLD bound—which itself is achievable through measurements on repeated preparations of the probe. On the other hand, for smaller values of $\lambda_1$, reaching the ultimate precision bound likely 
requires collective (entangled) measurements on multiple probe preparations.

We have also compared the optimized precision bounds of joint estimation with those of stepwise estimation methods, finding that joint estimation remains advantageous despite quantum noise induced by parameter incompatibility. Finally, we examined the precision achievable via general-dyne detection and found that it can approach the optimal precision in certain regimes.
\section*{Acknowledgment}
 PS acknowledges the UGC for the UGC Research Fellowship and the financial support from  Institution of Eminence (IoE), Banaras Hindu University, Varanasi, India for Grant under ``International Visiting Student Program". DKM acknowledges the financial support from the Science \& Engineering Research Board (SERB), New Delhi for CRG Grant (CRG/2021/005917) and Incentive Grant under, IoE, Banaras Hindu University, Varanasi, India. SO and MGAP acknowledge support from MUR - NextGenerationEU through Project 2022T25TR3 Recovering Information in Sloppy QUantum modEls (RISQUE). MGAP thank 
Massimo Frigerio, Jiayu He, Gabriele Fazio, Victor Montenegro, Chiranjib Mukhopadhyay, Abolfazl Bayat, Sara Dornetti, and Alessandro Ferraro for discussions.

\appendix

\section{Calculation for the joint measurement}

The explicit expressions of the matrix elements $Q_{jk}$ and $U_{jk}$, are given by 
\begin{align}
Q_{11}&= 16 \alpha^2+8 \\
Q_{12}&= Q_{21} = 16 \alpha ^2 \cos (2 \theta) \, \sinh (2 \lambda_1) \, \sin (2 \phi) +8 \left(2 \alpha ^2+1\right) \, \cos (2 \phi) \,, \\
Q_{22}&= 2 \Big[8 \alpha ^2 \sin (2 \theta)\, \sinh (4\lambda_1)\, \sin ^2 (2 \phi) +8 \alpha ^2 \cos (2 \theta) \, \sinh (2 \lambda_1)\, \sin (4 \phi)  \notag \\ 
&\hspace{0.5cm}-2 \left(4 \alpha ^2+1\right) \sinh ^2 (2\lambda_1)\, \cos (4 \phi) +(4 \alpha ^2+1) \cosh (4\lambda_1) +4 \alpha ^2+3\Big] \label{Qij}
\end{align}
and
\begin{align}
U_{11}&= U_{22} = 0\\
U_{12}&= - U_{21} = 8 \sin (2 \phi) \, \Big[2 \alpha ^2 \sin (2\theta) \,  \sinh (2\lambda_1) +\left(2 \alpha ^2+1\right) \cosh (2 \lambda_1) \Big]\,, \label{Uij}
\end{align}
respectively.

\bibliography{SqScra}

%apsrev4-2.bst 2019-01-14 (MD) hand-edited version of apsrev4-1.bst
%Control: key (0)
%Control: author (8) initials jnrlst
%Control: editor formatted (1) identically to author
%Control: production of article title (0) allowed
%Control: page (0) single
%Control: year (1) truncated
%Control: production of eprint (0) enabled
\begin{thebibliography}{46}%
\makeatletter
\providecommand \@ifxundefined [1]{%
 \@ifx{#1\undefined}
}%
\providecommand \@ifnum [1]{%
 \ifnum #1\expandafter \@firstoftwo
 \else \expandafter \@secondoftwo
 \fi
}%
\providecommand \@ifx [1]{%
 \ifx #1\expandafter \@firstoftwo
 \else \expandafter \@secondoftwo
 \fi
}%
\providecommand \natexlab [1]{#1}%
\providecommand \enquote  [1]{``#1''}%
\providecommand \bibnamefont  [1]{#1}%
\providecommand \bibfnamefont [1]{#1}%
\providecommand \citenamefont [1]{#1}%
\providecommand \href@noop [0]{\@secondoftwo}%
\providecommand \href [0]{\begingroup \@sanitize@url \@href}%
\providecommand \@href[1]{\@@startlink{#1}\@@href}%
\providecommand \@@href[1]{\endgroup#1\@@endlink}%
\providecommand \@sanitize@url [0]{\catcode `\\12\catcode `\$12\catcode
  `\&12\catcode `\#12\catcode `\^12\catcode `\_12\catcode `\%12\relax}%
\providecommand \@@startlink[1]{}%
\providecommand \@@endlink[0]{}%
\providecommand \url  [0]{\begingroup\@sanitize@url \@url }%
\providecommand \@url [1]{\endgroup\@href {#1}{\urlprefix }}%
\providecommand \urlprefix  [0]{URL }%
\providecommand \Eprint [0]{\href }%
\providecommand \doibase [0]{https://doi.org/}%
\providecommand \selectlanguage [0]{\@gobble}%
\providecommand \bibinfo  [0]{\@secondoftwo}%
\providecommand \bibfield  [0]{\@secondoftwo}%
\providecommand \translation [1]{[#1]}%
\providecommand \BibitemOpen [0]{}%
\providecommand \bibitemStop [0]{}%
\providecommand \bibitemNoStop [0]{.\EOS\space}%
\providecommand \EOS [0]{\spacefactor3000\relax}%
\providecommand \BibitemShut  [1]{\csname bibitem#1\endcsname}%
\let\auto@bib@innerbib\@empty
%</preamble>
\bibitem [{\citenamefont {Degen}\ \emph {et~al.}(2017)\citenamefont {Degen},
  \citenamefont {Reinhard},\ and\ \citenamefont
  {Cappellaro}}]{degen2017quantum}%
  \BibitemOpen
  \bibfield  {author} {\bibinfo {author} {\bibfnamefont {C.~L.}\ \bibnamefont
  {Degen}}, \bibinfo {author} {\bibfnamefont {F.}~\bibnamefont {Reinhard}},\
  and\ \bibinfo {author} {\bibfnamefont {P.}~\bibnamefont {Cappellaro}},\
  }\bibfield  {title} {\bibinfo {title} {Quantum sensing},\ }\href
  {https://doi.org/10.1103/RevModPhys.89.035002} {\bibfield  {journal}
  {\bibinfo  {journal} {Reviews of modern physics}\ }\textbf {\bibinfo {volume}
  {89}},\ \bibinfo {pages} {035002} (\bibinfo {year} {2017})}\BibitemShut
  {NoStop}%
\bibitem [{\citenamefont {Montenegro}\ \emph {et~al.}(2025)\citenamefont
  {Montenegro}, \citenamefont {Mukhopadhyay}, \citenamefont {Yousefjani},
  \citenamefont {Sarkar}, \citenamefont {Mishra}, \citenamefont {Paris},\ and\
  \citenamefont {Bayat}}]{montenegro2025quantum}%
  \BibitemOpen
  \bibfield  {author} {\bibinfo {author} {\bibfnamefont {V.}~\bibnamefont
  {Montenegro}}, \bibinfo {author} {\bibfnamefont {C.}~\bibnamefont
  {Mukhopadhyay}}, \bibinfo {author} {\bibfnamefont {R.}~\bibnamefont
  {Yousefjani}}, \bibinfo {author} {\bibfnamefont {S.}~\bibnamefont {Sarkar}},
  \bibinfo {author} {\bibfnamefont {U.}~\bibnamefont {Mishra}}, \bibinfo
  {author} {\bibfnamefont {M.~G.~A.}\ \bibnamefont {Paris}},\ and\ \bibinfo
  {author} {\bibfnamefont {A.}~\bibnamefont {Bayat}},\ }\bibfield  {title}
  {\bibinfo {title} {Quantum metrology and sensing with many-body systems},\
  }\href {https://doi.org/https://doi.org/10.1016/j.physrep.2025.05.005}
  {\bibfield  {journal} {\bibinfo  {journal} {Physics Reports}\ }\textbf
  {\bibinfo {volume} {1134}},\ \bibinfo {pages} {1} (\bibinfo {year}
  {2025})}\BibitemShut {NoStop}%
\bibitem [{\citenamefont {Helstrom}(1976)}]{1976Helstrom}%
  \BibitemOpen
  \bibfield  {author} {\bibinfo {author} {\bibfnamefont {C.~W.}\ \bibnamefont
  {Helstrom}},\ }\href {https://books.google.co.in/books?id=Ne3iT\_QLcsMC}
  {\emph {\bibinfo {title} {Quantum {D}etection and {E}stimation {T}heory}}}\
  (\bibinfo  {publisher} {Academic Press, San Diego, CA},\ \bibinfo {year}
  {1976})\BibitemShut {NoStop}%
\bibitem [{\citenamefont {Holevo}(2011)}]{Holevo}%
  \BibitemOpen
  \bibfield  {author} {\bibinfo {author} {\bibfnamefont {A.}~\bibnamefont
  {Holevo}},\ }\href
  {https://doi.org/https://doi.org/10.1007/978-88-7642-378-9} {\emph {\bibinfo
  {title} {Probabilistic and Statistical Aspects of Quantum Theory}}}\
  (\bibinfo  {publisher} {Edizioni della Normale Pisa},\ \bibinfo {year}
  {2011})\BibitemShut {NoStop}%
\bibitem [{\citenamefont {Rehacek}\ and\ \citenamefont {Paris}(2004)}]{MGAP}%
  \BibitemOpen
  \bibfield  {author} {\bibinfo {author} {\bibfnamefont {J.}~\bibnamefont
  {Rehacek}}\ and\ \bibinfo {author} {\bibfnamefont {M.~G.~A.}\ \bibnamefont
  {Paris}},\ }\href {https://doi.org/https://doi.org/10.1007/b98673} {\emph
  {\bibinfo {title} {Quantum State Estimation}}}\ (\bibinfo  {publisher}
  {Springer Berlin, Heidelberg},\ \bibinfo {year} {2004})\BibitemShut {NoStop}%
\bibitem [{\citenamefont {Paris}(2009)}]{doi:10.1142/S0219749909004839}%
  \BibitemOpen
  \bibfield  {author} {\bibinfo {author} {\bibfnamefont {M.~G.~A.}\
  \bibnamefont {Paris}},\ }\bibfield  {title} {\bibinfo {title} {Quantum
  estimation for quantum technology},\ }\href
  {https://doi.org/10.1142/S0219749909004839} {\bibfield  {journal} {\bibinfo
  {journal} {International Journal of Quantum Information}\ }\textbf {\bibinfo
  {volume} {07}},\ \bibinfo {pages} {125} (\bibinfo {year} {2009})}\BibitemShut
  {NoStop}%
\bibitem [{\citenamefont {Drummond}\ and\ \citenamefont
  {Ficek}(2013)}]{drummond2013quantum}%
  \BibitemOpen
  \bibfield  {author} {\bibinfo {author} {\bibfnamefont {P.~D.}\ \bibnamefont
  {Drummond}}\ and\ \bibinfo {author} {\bibfnamefont {Z.}~\bibnamefont
  {Ficek}},\ }\href {https://doi.org/https://doi.org/10.1007/978-3-662-09645-1}
  {\emph {\bibinfo {title} {Quantum squeezing}}},\ Vol.~\bibinfo {volume} {27}\
  (\bibinfo  {publisher} {Springer Science \& Business Media},\ \bibinfo {year}
  {2013})\BibitemShut {NoStop}%
\bibitem [{\citenamefont {Slusher}\ \emph {et~al.}(1985)\citenamefont
  {Slusher}, \citenamefont {Hollberg}, \citenamefont {Yurke}, \citenamefont
  {Mertz},\ and\ \citenamefont {Valley}}]{PhysRevLett.55.2409}%
  \BibitemOpen
  \bibfield  {author} {\bibinfo {author} {\bibfnamefont {R.~E.}\ \bibnamefont
  {Slusher}}, \bibinfo {author} {\bibfnamefont {L.~W.}\ \bibnamefont
  {Hollberg}}, \bibinfo {author} {\bibfnamefont {B.}~\bibnamefont {Yurke}},
  \bibinfo {author} {\bibfnamefont {J.~C.}\ \bibnamefont {Mertz}},\ and\
  \bibinfo {author} {\bibfnamefont {J.~F.}\ \bibnamefont {Valley}},\ }\bibfield
   {title} {\bibinfo {title} {Observation of squeezed states generated by
  four-wave mixing in an optical cavity},\ }\href
  {https://doi.org/10.1103/PhysRevLett.55.2409} {\bibfield  {journal} {\bibinfo
   {journal} {Phys. Rev. Lett.}\ }\textbf {\bibinfo {volume} {55}},\ \bibinfo
  {pages} {2409} (\bibinfo {year} {1985})}\BibitemShut {NoStop}%
\bibitem [{\citenamefont {D'Ariano}\ \emph {et~al.}(2001)\citenamefont
  {D'Ariano}, \citenamefont {Lo~Presti},\ and\ \citenamefont
  {Paris}}]{PhysRevLett.87.270404}%
  \BibitemOpen
  \bibfield  {author} {\bibinfo {author} {\bibfnamefont {G.~M.}\ \bibnamefont
  {D'Ariano}}, \bibinfo {author} {\bibfnamefont {P.}~\bibnamefont
  {Lo~Presti}},\ and\ \bibinfo {author} {\bibfnamefont {M.~G.~A.}\ \bibnamefont
  {Paris}},\ }\bibfield  {title} {\bibinfo {title} {Using entanglement improves
  the precision of quantum measurements},\ }\href
  {https://doi.org/10.1103/PhysRevLett.87.270404} {\bibfield  {journal}
  {\bibinfo  {journal} {Phys. Rev. Lett.}\ }\textbf {\bibinfo {volume} {87}},\
  \bibinfo {pages} {270404} (\bibinfo {year} {2001})}\BibitemShut {NoStop}%
\bibitem [{\citenamefont {Braunstein}\ and\ \citenamefont
  {Caves}(1994)}]{PhysRevLett.72.3439}%
  \BibitemOpen
  \bibfield  {author} {\bibinfo {author} {\bibfnamefont {S.~L.}\ \bibnamefont
  {Braunstein}}\ and\ \bibinfo {author} {\bibfnamefont {C.~M.}\ \bibnamefont
  {Caves}},\ }\bibfield  {title} {\bibinfo {title} {Statistical distance and
  the geometry of quantum states},\ }\href
  {https://doi.org/10.1103/PhysRevLett.72.3439} {\bibfield  {journal} {\bibinfo
   {journal} {Phys. Rev. Lett.}\ }\textbf {\bibinfo {volume} {72}},\ \bibinfo
  {pages} {3439} (\bibinfo {year} {1994})}\BibitemShut {NoStop}%
\bibitem [{\citenamefont {Sharma}\ \emph {et~al.}(2025)\citenamefont {Sharma},
  \citenamefont {Kumar}, \citenamefont {Shukla},\ and\ \citenamefont
  {Mishra}}]{SHARMA2025130459}%
  \BibitemOpen
  \bibfield  {author} {\bibinfo {author} {\bibfnamefont {P.}~\bibnamefont
  {Sharma}}, \bibinfo {author} {\bibfnamefont {A.}~\bibnamefont {Kumar}},
  \bibinfo {author} {\bibfnamefont {G.}~\bibnamefont {Shukla}},\ and\ \bibinfo
  {author} {\bibfnamefont {D.~K.}\ \bibnamefont {Mishra}},\ }\bibfield  {title}
  {\bibinfo {title} {Optimizing phase sensitivity of mach-zehnder
  interferometer having superposition of coherent state with
  single-photon-added coherent state},\ }\href
  {https://doi.org/https://doi.org/10.1016/j.physleta.2025.130459} {\bibfield
  {journal} {\bibinfo  {journal} {Physics Letters A}\ }\textbf {\bibinfo
  {volume} {543}},\ \bibinfo {pages} {130459} (\bibinfo {year}
  {2025})}\BibitemShut {NoStop}%
\bibitem [{\citenamefont {Rossi}\ and\ \citenamefont
  {Paris}(2015)}]{rossi2015entangled}%
  \BibitemOpen
  \bibfield  {author} {\bibinfo {author} {\bibfnamefont {M.~A.~C.}\
  \bibnamefont {Rossi}}\ and\ \bibinfo {author} {\bibfnamefont {M.~G.~A.}\
  \bibnamefont {Paris}},\ }\bibfield  {title} {\bibinfo {title} {Entangled
  quantum probes for dynamical environmental noise},\ }\href
  {https://doi.org/10.1103/PhysRevA.92.010302} {\bibfield  {journal} {\bibinfo
  {journal} {Physical Review A}\ }\textbf {\bibinfo {volume} {92}},\ \bibinfo
  {pages} {010302} (\bibinfo {year} {2015})}\BibitemShut {NoStop}%
\bibitem [{\citenamefont {Bina}\ \emph {et~al.}(2018)\citenamefont {Bina},
  \citenamefont {Grasselli},\ and\ \citenamefont {Paris}}]{bina2018continuous}%
  \BibitemOpen
  \bibfield  {author} {\bibinfo {author} {\bibfnamefont {M.}~\bibnamefont
  {Bina}}, \bibinfo {author} {\bibfnamefont {F.}~\bibnamefont {Grasselli}},\
  and\ \bibinfo {author} {\bibfnamefont {M.~G.~A.}\ \bibnamefont {Paris}},\
  }\bibfield  {title} {\bibinfo {title} {Continuous-variable quantum probes for
  structured environments},\ }\href
  {https://doi.org/10.1103/PhysRevA.97.012125} {\bibfield  {journal} {\bibinfo
  {journal} {Physical Review A}\ }\textbf {\bibinfo {volume} {97}},\ \bibinfo
  {pages} {012125} (\bibinfo {year} {2018})}\BibitemShut {NoStop}%
\bibitem [{\citenamefont {Fiderer}\ \emph {et~al.}(2021)\citenamefont
  {Fiderer}, \citenamefont {Tufarelli}, \citenamefont {Piano},\ and\
  \citenamefont {Adesso}}]{PRXQuantum.2.020308}%
  \BibitemOpen
  \bibfield  {author} {\bibinfo {author} {\bibfnamefont {L.~J.}\ \bibnamefont
  {Fiderer}}, \bibinfo {author} {\bibfnamefont {T.}~\bibnamefont {Tufarelli}},
  \bibinfo {author} {\bibfnamefont {S.}~\bibnamefont {Piano}},\ and\ \bibinfo
  {author} {\bibfnamefont {G.}~\bibnamefont {Adesso}},\ }\bibfield  {title}
  {\bibinfo {title} {General expressions for the quantum fisher information
  matrix with applications to discrete quantum imaging},\ }\href
  {https://doi.org/10.1103/PRXQuantum.2.020308} {\bibfield  {journal} {\bibinfo
   {journal} {PRX Quantum}\ }\textbf {\bibinfo {volume} {2}},\ \bibinfo {pages}
  {020308} (\bibinfo {year} {2021})}\BibitemShut {NoStop}%
\bibitem [{\citenamefont {Sharma}\ \emph {et~al.}(2024)\citenamefont {Sharma},
  \citenamefont {Pandey}, \citenamefont {Shukla},\ and\ \citenamefont
  {Mishra}}]{SHARMA24OC}%
  \BibitemOpen
  \bibfield  {author} {\bibinfo {author} {\bibfnamefont {P.}~\bibnamefont
  {Sharma}}, \bibinfo {author} {\bibfnamefont {A.~K.}\ \bibnamefont {Pandey}},
  \bibinfo {author} {\bibfnamefont {G.}~\bibnamefont {Shukla}},\ and\ \bibinfo
  {author} {\bibfnamefont {D.~K.}\ \bibnamefont {Mishra}},\ }\bibfield  {title}
  {\bibinfo {title} {Enhancement in phase sensitivity of {SU}(1,1)
  interferometer with kerr state seeding},\ }\href
  {https://doi.org/https://doi.org/10.1016/j.optcom.2024.131028} {\bibfield
  {journal} {\bibinfo  {journal} {Optics Communications}\ }\textbf {\bibinfo
  {volume} {573}},\ \bibinfo {pages} {131028} (\bibinfo {year}
  {2024})}\BibitemShut {NoStop}%
\bibitem [{\citenamefont {Giovannetti}\ \emph {et~al.}(2004)\citenamefont
  {Giovannetti}, \citenamefont {Lloyd},\ and\ \citenamefont
  {Maccone}}]{SethLloyd}%
  \BibitemOpen
  \bibfield  {author} {\bibinfo {author} {\bibfnamefont {V.}~\bibnamefont
  {Giovannetti}}, \bibinfo {author} {\bibfnamefont {S.}~\bibnamefont {Lloyd}},\
  and\ \bibinfo {author} {\bibfnamefont {L.}~\bibnamefont {Maccone}},\
  }\bibfield  {title} {\bibinfo {title} {Quantum-enhanced measurements: Beating
  the standard quantum limit},\ }\href
  {https://doi.org/10.1126/science.1104149} {\bibfield  {journal} {\bibinfo
  {journal} {Science}\ }\textbf {\bibinfo {volume} {306}},\ \bibinfo {pages}
  {1330} (\bibinfo {year} {2004})}\BibitemShut {NoStop}%
\bibitem [{\citenamefont {Albarelli}\ \emph {et~al.}(2020)\citenamefont
  {Albarelli}, \citenamefont {Barbieri}, \citenamefont {Genoni},\ and\
  \citenamefont {Gianani}}]{ALBARELLI2020126311}%
  \BibitemOpen
  \bibfield  {author} {\bibinfo {author} {\bibfnamefont {F.}~\bibnamefont
  {Albarelli}}, \bibinfo {author} {\bibfnamefont {M.}~\bibnamefont {Barbieri}},
  \bibinfo {author} {\bibfnamefont {M.~G.}\ \bibnamefont {Genoni}},\ and\
  \bibinfo {author} {\bibfnamefont {I.}~\bibnamefont {Gianani}},\ }\bibfield
  {title} {\bibinfo {title} {A perspective on multiparameter quantum metrology:
  From theoretical tools to applications in quantum imaging},\ }\href
  {https://doi.org/https://doi.org/10.1016/j.physleta.2020.126311} {\bibfield
  {journal} {\bibinfo  {journal} {Physics Letters A}\ }\textbf {\bibinfo
  {volume} {384}},\ \bibinfo {pages} {126311} (\bibinfo {year}
  {2020})}\BibitemShut {NoStop}%
\bibitem [{\citenamefont {Liu}\ \emph {et~al.}(2020)\citenamefont {Liu},
  \citenamefont {Yuan}, \citenamefont {Lu},\ and\ \citenamefont
  {Wang}}]{liu2020quantum}%
  \BibitemOpen
  \bibfield  {author} {\bibinfo {author} {\bibfnamefont {J.}~\bibnamefont
  {Liu}}, \bibinfo {author} {\bibfnamefont {H.}~\bibnamefont {Yuan}}, \bibinfo
  {author} {\bibfnamefont {X.-M.}\ \bibnamefont {Lu}},\ and\ \bibinfo {author}
  {\bibfnamefont {X.}~\bibnamefont {Wang}},\ }\bibfield  {title} {\bibinfo
  {title} {Quantum fisher information matrix and multiparameter estimation},\
  }\href {https://doi.org/10.1088/1751-8121/ab5d4d} {\bibfield  {journal}
  {\bibinfo  {journal} {Journal of Physics A: Mathematical and Theoretical}\
  }\textbf {\bibinfo {volume} {53}},\ \bibinfo {pages} {023001} (\bibinfo
  {year} {2020})}\BibitemShut {NoStop}%
\bibitem [{\citenamefont {Tsang}\ \emph {et~al.}(2016)\citenamefont {Tsang},
  \citenamefont {Nair},\ and\ \citenamefont {Lu}}]{PhysRevX.6.031033}%
  \BibitemOpen
  \bibfield  {author} {\bibinfo {author} {\bibfnamefont {M.}~\bibnamefont
  {Tsang}}, \bibinfo {author} {\bibfnamefont {R.}~\bibnamefont {Nair}},\ and\
  \bibinfo {author} {\bibfnamefont {X.-M.}\ \bibnamefont {Lu}},\ }\bibfield
  {title} {\bibinfo {title} {Quantum theory of superresolution for two
  incoherent optical point sources},\ }\href
  {https://doi.org/10.1103/PhysRevX.6.031033} {\bibfield  {journal} {\bibinfo
  {journal} {Phys. Rev. X}\ }\textbf {\bibinfo {volume} {6}},\ \bibinfo {pages}
  {031033} (\bibinfo {year} {2016})}\BibitemShut {NoStop}%
\bibitem [{\citenamefont {Taylor}\ \emph {et~al.}(2008)\citenamefont {Taylor},
  \citenamefont {Cappellaro}, \citenamefont {Childress}, \citenamefont {Jiang},
  \citenamefont {Budker}, \citenamefont {Hemmer}, \citenamefont {Yacoby},
  \citenamefont {Walsworth},\ and\ \citenamefont {Lukin}}]{taylor2008high}%
  \BibitemOpen
  \bibfield  {author} {\bibinfo {author} {\bibfnamefont {J.~M.}\ \bibnamefont
  {Taylor}}, \bibinfo {author} {\bibfnamefont {P.}~\bibnamefont {Cappellaro}},
  \bibinfo {author} {\bibfnamefont {L.}~\bibnamefont {Childress}}, \bibinfo
  {author} {\bibfnamefont {L.}~\bibnamefont {Jiang}}, \bibinfo {author}
  {\bibfnamefont {D.}~\bibnamefont {Budker}}, \bibinfo {author} {\bibfnamefont
  {P.~R.}\ \bibnamefont {Hemmer}}, \bibinfo {author} {\bibfnamefont
  {A.}~\bibnamefont {Yacoby}}, \bibinfo {author} {\bibfnamefont
  {R.}~\bibnamefont {Walsworth}},\ and\ \bibinfo {author} {\bibfnamefont
  {M.~D.}\ \bibnamefont {Lukin}},\ }\bibfield  {title} {\bibinfo {title}
  {High-sensitivity diamond magnetometer with nanoscale resolution},\ }\href
  {https://doi.org/10.1038/nphys1075} {\bibfield  {journal} {\bibinfo
  {journal} {Nature Physics}\ }\textbf {\bibinfo {volume} {4}},\ \bibinfo
  {pages} {810} (\bibinfo {year} {2008})}\BibitemShut {NoStop}%
\bibitem [{\citenamefont {Baumgratz}\ and\ \citenamefont
  {Datta}(2016)}]{PhysRevLett.116.030801}%
  \BibitemOpen
  \bibfield  {author} {\bibinfo {author} {\bibfnamefont {T.}~\bibnamefont
  {Baumgratz}}\ and\ \bibinfo {author} {\bibfnamefont {A.}~\bibnamefont
  {Datta}},\ }\bibfield  {title} {\bibinfo {title} {Quantum enhanced estimation
  of a multidimensional field},\ }\href
  {https://doi.org/10.1103/PhysRevLett.116.030801} {\bibfield  {journal}
  {\bibinfo  {journal} {Phys. Rev. Lett.}\ }\textbf {\bibinfo {volume} {116}},\
  \bibinfo {pages} {030801} (\bibinfo {year} {2016})}\BibitemShut {NoStop}%
\bibitem [{\citenamefont {Humphreys}\ \emph {et~al.}(2013)\citenamefont
  {Humphreys}, \citenamefont {Barbieri}, \citenamefont {Datta},\ and\
  \citenamefont {Walmsley}}]{PhysRevLett.111.070403}%
  \BibitemOpen
  \bibfield  {author} {\bibinfo {author} {\bibfnamefont {P.~C.}\ \bibnamefont
  {Humphreys}}, \bibinfo {author} {\bibfnamefont {M.}~\bibnamefont {Barbieri}},
  \bibinfo {author} {\bibfnamefont {A.}~\bibnamefont {Datta}},\ and\ \bibinfo
  {author} {\bibfnamefont {I.~A.}\ \bibnamefont {Walmsley}},\ }\bibfield
  {title} {\bibinfo {title} {Quantum enhanced multiple phase estimation},\
  }\href {https://doi.org/10.1103/PhysRevLett.111.070403} {\bibfield  {journal}
  {\bibinfo  {journal} {Phys. Rev. Lett.}\ }\textbf {\bibinfo {volume} {111}},\
  \bibinfo {pages} {070403} (\bibinfo {year} {2013})}\BibitemShut {NoStop}%
\bibitem [{\citenamefont {Proctor}\ \emph {et~al.}(2018)\citenamefont
  {Proctor}, \citenamefont {Knott},\ and\ \citenamefont
  {Dunningham}}]{PhysRevLett.120.080501}%
  \BibitemOpen
  \bibfield  {author} {\bibinfo {author} {\bibfnamefont {T.~J.}\ \bibnamefont
  {Proctor}}, \bibinfo {author} {\bibfnamefont {P.~A.}\ \bibnamefont {Knott}},\
  and\ \bibinfo {author} {\bibfnamefont {J.~A.}\ \bibnamefont {Dunningham}},\
  }\bibfield  {title} {\bibinfo {title} {Multiparameter estimation in networked
  quantum sensors},\ }\href {https://doi.org/10.1103/PhysRevLett.120.080501}
  {\bibfield  {journal} {\bibinfo  {journal} {Phys. Rev. Lett.}\ }\textbf
  {\bibinfo {volume} {120}},\ \bibinfo {pages} {080501} (\bibinfo {year}
  {2018})}\BibitemShut {NoStop}%
\bibitem [{\citenamefont {Gill}\ and\ \citenamefont
  {Massar}(2000)}]{PhysRevA.61.042312}%
  \BibitemOpen
  \bibfield  {author} {\bibinfo {author} {\bibfnamefont {R.~D.}\ \bibnamefont
  {Gill}}\ and\ \bibinfo {author} {\bibfnamefont {S.}~\bibnamefont {Massar}},\
  }\bibfield  {title} {\bibinfo {title} {State estimation for large
  ensembles},\ }\href {https://doi.org/10.1103/PhysRevA.61.042312} {\bibfield
  {journal} {\bibinfo  {journal} {Phys. Rev. A}\ }\textbf {\bibinfo {volume}
  {61}},\ \bibinfo {pages} {042312} (\bibinfo {year} {2000})}\BibitemShut
  {NoStop}%
\bibitem [{\citenamefont {Ragy}\ \emph {et~al.}(2016)\citenamefont {Ragy},
  \citenamefont {Jarzyna},\ and\ \citenamefont
  {Demkowicz-Dobrza\ifmmode~\acute{n}\else
  \'{n}\fi{}ski}}]{PhysRevA.94.052108}%
  \BibitemOpen
  \bibfield  {author} {\bibinfo {author} {\bibfnamefont {S.}~\bibnamefont
  {Ragy}}, \bibinfo {author} {\bibfnamefont {M.}~\bibnamefont {Jarzyna}},\ and\
  \bibinfo {author} {\bibfnamefont {R.}~\bibnamefont
  {Demkowicz-Dobrza\ifmmode~\acute{n}\else \'{n}\fi{}ski}},\ }\bibfield
  {title} {\bibinfo {title} {Compatibility in multiparameter quantum
  metrology},\ }\href {https://doi.org/10.1103/PhysRevA.94.052108} {\bibfield
  {journal} {\bibinfo  {journal} {Phys. Rev. A}\ }\textbf {\bibinfo {volume}
  {94}},\ \bibinfo {pages} {052108} (\bibinfo {year} {2016})}\BibitemShut
  {NoStop}%
\bibitem [{\citenamefont {Szczykulska}\ \emph {et~al.}(2016)\citenamefont
  {Szczykulska}, \citenamefont {Baumgratz},\ and\ \citenamefont
  {and}}]{Szczykulska03072016}%
  \BibitemOpen
  \bibfield  {author} {\bibinfo {author} {\bibfnamefont {M.}~\bibnamefont
  {Szczykulska}}, \bibinfo {author} {\bibfnamefont {T.}~\bibnamefont
  {Baumgratz}},\ and\ \bibinfo {author} {\bibfnamefont {A.~D.}\ \bibnamefont
  {and}},\ }\bibfield  {title} {\bibinfo {title} {Multi-parameter quantum
  metrology},\ }\href {https://doi.org/10.1080/23746149.2016.1230476}
  {\bibfield  {journal} {\bibinfo  {journal} {Advances in Physics: X}\ }\textbf
  {\bibinfo {volume} {1}},\ \bibinfo {pages} {621} (\bibinfo {year}
  {2016})}\BibitemShut {NoStop}%
\bibitem [{\citenamefont {Suzuki}(2016)}]{suzuki2016explicit}%
  \BibitemOpen
  \bibfield  {author} {\bibinfo {author} {\bibfnamefont {J.}~\bibnamefont
  {Suzuki}},\ }\bibfield  {title} {\bibinfo {title} {Explicit formula for the
  holevo bound for two-parameter qubit-state estimation problem},\ }\bibfield
  {journal} {\bibinfo  {journal} {Journal of Mathematical Physics}\ }\textbf
  {\bibinfo {volume} {57}},\ \href {https://doi.org/10.1063/1.4945086}
  {10.1063/1.4945086} (\bibinfo {year} {2016})\BibitemShut {NoStop}%
\bibitem [{\citenamefont {Albarelli}\ \emph {et~al.}(2019)\citenamefont
  {Albarelli}, \citenamefont {Friel},\ and\ \citenamefont
  {Datta}}]{PhysRevLett.123.200503}%
  \BibitemOpen
  \bibfield  {author} {\bibinfo {author} {\bibfnamefont {F.}~\bibnamefont
  {Albarelli}}, \bibinfo {author} {\bibfnamefont {J.~F.}\ \bibnamefont
  {Friel}},\ and\ \bibinfo {author} {\bibfnamefont {A.}~\bibnamefont {Datta}},\
  }\bibfield  {title} {\bibinfo {title} {Evaluating the holevo cram\'er-rao
  bound for multiparameter quantum metrology},\ }\href
  {https://doi.org/10.1103/PhysRevLett.123.200503} {\bibfield  {journal}
  {\bibinfo  {journal} {Phys. Rev. Lett.}\ }\textbf {\bibinfo {volume} {123}},\
  \bibinfo {pages} {200503} (\bibinfo {year} {2019})}\BibitemShut {NoStop}%
\bibitem [{\citenamefont {Demkowicz-Dobrzański}\ \emph
  {et~al.}(2020)\citenamefont {Demkowicz-Dobrzański}, \citenamefont
  {Górecki},\ and\ \citenamefont {Guţă}}]{Demkowicz20}%
  \BibitemOpen
  \bibfield  {author} {\bibinfo {author} {\bibfnamefont {R.}~\bibnamefont
  {Demkowicz-Dobrzański}}, \bibinfo {author} {\bibfnamefont {W.}~\bibnamefont
  {Górecki}},\ and\ \bibinfo {author} {\bibfnamefont {M.}~\bibnamefont
  {Guţă}},\ }\bibfield  {title} {\bibinfo {title} {Multi-parameter estimation
  beyond quantum fisher information},\ }\href
  {https://doi.org/10.1088/1751-8121/ab8ef3} {\bibfield  {journal} {\bibinfo
  {journal} {Journal of Physics A: Mathematical and Theoretical}\ }\textbf
  {\bibinfo {volume} {53}},\ \bibinfo {pages} {363001} (\bibinfo {year}
  {2020})}\BibitemShut {NoStop}%
\bibitem [{\citenamefont {Rubio}\ and\ \citenamefont
  {Dunningham}(2020)}]{RubioPhysRevA.101.032114}%
  \BibitemOpen
  \bibfield  {author} {\bibinfo {author} {\bibfnamefont {J.}~\bibnamefont
  {Rubio}}\ and\ \bibinfo {author} {\bibfnamefont {J.}~\bibnamefont
  {Dunningham}},\ }\bibfield  {title} {\bibinfo {title} {Bayesian
  multiparameter quantum metrology with limited data},\ }\href
  {https://doi.org/10.1103/PhysRevA.101.032114} {\bibfield  {journal} {\bibinfo
   {journal} {Phys. Rev. A}\ }\textbf {\bibinfo {volume} {101}},\ \bibinfo
  {pages} {032114} (\bibinfo {year} {2020})}\BibitemShut {NoStop}%
\bibitem [{\citenamefont {Lee}\ \emph {et~al.}(2023)\citenamefont {Lee},
  \citenamefont {Gagatsos}, \citenamefont {Guha},\ and\ \citenamefont
  {Ashok}}]{Lee9919344}%
  \BibitemOpen
  \bibfield  {author} {\bibinfo {author} {\bibfnamefont {K.~K.}\ \bibnamefont
  {Lee}}, \bibinfo {author} {\bibfnamefont {C.~N.}\ \bibnamefont {Gagatsos}},
  \bibinfo {author} {\bibfnamefont {S.}~\bibnamefont {Guha}},\ and\ \bibinfo
  {author} {\bibfnamefont {A.}~\bibnamefont {Ashok}},\ }\bibfield  {title}
  {\bibinfo {title} {Quantum-inspired multi-parameter adaptive bayesian
  estimation for sensing and imaging},\ }\href
  {https://doi.org/10.1109/JSTSP.2022.3214774} {\bibfield  {journal} {\bibinfo
  {journal} {IEEE Journal of Selected Topics in Signal Processing}\ }\textbf
  {\bibinfo {volume} {17}},\ \bibinfo {pages} {491} (\bibinfo {year}
  {2023})}\BibitemShut {NoStop}%
\bibitem [{\citenamefont {Razavian}\ \emph {et~al.}(2020)\citenamefont
  {Razavian}, \citenamefont {Paris},\ and\ \citenamefont
  {Genoni}}]{razavian2020quantumness}%
  \BibitemOpen
  \bibfield  {author} {\bibinfo {author} {\bibfnamefont {S.}~\bibnamefont
  {Razavian}}, \bibinfo {author} {\bibfnamefont {M.~G.~A.}\ \bibnamefont
  {Paris}},\ and\ \bibinfo {author} {\bibfnamefont {M.~G.}\ \bibnamefont
  {Genoni}},\ }\bibfield  {title} {\bibinfo {title} {On the quantumness of
  multiparameter estimation problems for qubit systems},\ }\href
  {https://doi.org/10.3390/e22111197} {\bibfield  {journal} {\bibinfo
  {journal} {Entropy}\ }\textbf {\bibinfo {volume} {22}},\ \bibinfo {pages}
  {1197} (\bibinfo {year} {2020})}\BibitemShut {NoStop}%
\bibitem [{\citenamefont {Bressanini}\ \emph {et~al.}(2024)\citenamefont
  {Bressanini}, \citenamefont {Genoni}, \citenamefont {Kim},\ and\
  \citenamefont {Paris}}]{Bressanini_2024}%
  \BibitemOpen
  \bibfield  {author} {\bibinfo {author} {\bibfnamefont {G.}~\bibnamefont
  {Bressanini}}, \bibinfo {author} {\bibfnamefont {M.~G.}\ \bibnamefont
  {Genoni}}, \bibinfo {author} {\bibfnamefont {M.~S.}\ \bibnamefont {Kim}},\
  and\ \bibinfo {author} {\bibfnamefont {M.~G.~A.}\ \bibnamefont {Paris}},\
  }\bibfield  {title} {\bibinfo {title} {Multi-parameter quantum estimation of
  single- and two-mode pure gaussian states},\ }\href
  {https://doi.org/10.1088/1751-8121/ad6364} {\bibfield  {journal} {\bibinfo
  {journal} {Journal of Physics A: Mathematical and Theoretical}\ }\textbf
  {\bibinfo {volume} {57}},\ \bibinfo {pages} {315305} (\bibinfo {year}
  {2024})}\BibitemShut {NoStop}%
\bibitem [{\citenamefont {Frigerio}\ and\ \citenamefont
  {Paris}(2024)}]{frigerio2024overcomingsloppinessenhancedmetrology}%
  \BibitemOpen
  \bibfield  {author} {\bibinfo {author} {\bibfnamefont {M.}~\bibnamefont
  {Frigerio}}\ and\ \bibinfo {author} {\bibfnamefont {M.~G.~A.}\ \bibnamefont
  {Paris}},\ }\href {https://arxiv.org/abs/2410.02989} {\bibinfo {title}
  {Overcoming sloppiness for enhanced metrology in continuous-variable quantum
  statistical models}} (\bibinfo {year} {2024}),\ \Eprint
  {https://arxiv.org/abs/2410.02989} {arXiv:2410.02989 [quant-ph]} \BibitemShut
  {NoStop}%
\bibitem [{\citenamefont {Gessner}\ \emph {et~al.}(2018)\citenamefont
  {Gessner}, \citenamefont {Pezz\`e},\ and\ \citenamefont
  {Smerzi}}]{PhysRevLett.121.130503}%
  \BibitemOpen
  \bibfield  {author} {\bibinfo {author} {\bibfnamefont {M.}~\bibnamefont
  {Gessner}}, \bibinfo {author} {\bibfnamefont {L.}~\bibnamefont {Pezz\`e}},\
  and\ \bibinfo {author} {\bibfnamefont {A.}~\bibnamefont {Smerzi}},\
  }\bibfield  {title} {\bibinfo {title} {Sensitivity bounds for multiparameter
  quantum metrology},\ }\href {https://doi.org/10.1103/PhysRevLett.121.130503}
  {\bibfield  {journal} {\bibinfo  {journal} {Phys. Rev. Lett.}\ }\textbf
  {\bibinfo {volume} {121}},\ \bibinfo {pages} {130503} (\bibinfo {year}
  {2018})}\BibitemShut {NoStop}%
\bibitem [{\citenamefont {Candeloro}\ \emph {et~al.}(2024)\citenamefont
  {Candeloro}, \citenamefont {Pazhotan},\ and\ \citenamefont
  {Paris}}]{Candeloro_2024}%
  \BibitemOpen
  \bibfield  {author} {\bibinfo {author} {\bibfnamefont {A.}~\bibnamefont
  {Candeloro}}, \bibinfo {author} {\bibfnamefont {Z.}~\bibnamefont
  {Pazhotan}},\ and\ \bibinfo {author} {\bibfnamefont {M.~G.~A.}\ \bibnamefont
  {Paris}},\ }\bibfield  {title} {\bibinfo {title} {Dimension matters:
  precision and incompatibility in multi-parameter quantum estimation models},\
  }\href {https://doi.org/10.1088/2058-9565/ad7498} {\bibfield  {journal}
  {\bibinfo  {journal} {Quantum Science and Technology}\ }\textbf {\bibinfo
  {volume} {9}},\ \bibinfo {pages} {045045} (\bibinfo {year}
  {2024})}\BibitemShut {NoStop}%
\bibitem [{\citenamefont {Milburn}\ \emph {et~al.}(1994)\citenamefont
  {Milburn}, \citenamefont {Chen},\ and\ \citenamefont
  {Jones}}]{PhysRevA.50.801}%
  \BibitemOpen
  \bibfield  {author} {\bibinfo {author} {\bibfnamefont {G.~J.}\ \bibnamefont
  {Milburn}}, \bibinfo {author} {\bibfnamefont {W.-Y.}\ \bibnamefont {Chen}},\
  and\ \bibinfo {author} {\bibfnamefont {K.~R.}\ \bibnamefont {Jones}},\
  }\bibfield  {title} {\bibinfo {title} {Hyperbolic phase and squeeze-parameter
  estimation},\ }\href {https://doi.org/10.1103/PhysRevA.50.801} {\bibfield
  {journal} {\bibinfo  {journal} {Phys. Rev. A}\ }\textbf {\bibinfo {volume}
  {50}},\ \bibinfo {pages} {801} (\bibinfo {year} {1994})}\BibitemShut
  {NoStop}%
\bibitem [{\citenamefont {Bollini}\ and\ \citenamefont
  {Oxman}(1993)}]{PhysRevA.47.2339}%
  \BibitemOpen
  \bibfield  {author} {\bibinfo {author} {\bibfnamefont {C.~G.}\ \bibnamefont
  {Bollini}}\ and\ \bibinfo {author} {\bibfnamefont {L.~E.}\ \bibnamefont
  {Oxman}},\ }\bibfield  {title} {\bibinfo {title} {Shannon entropy and the
  eigenstates of the single-mode squeeze operator},\ }\href
  {https://doi.org/10.1103/PhysRevA.47.2339} {\bibfield  {journal} {\bibinfo
  {journal} {Phys. Rev. A}\ }\textbf {\bibinfo {volume} {47}},\ \bibinfo
  {pages} {2339} (\bibinfo {year} {1993})}\BibitemShut {NoStop}%
\bibitem [{\citenamefont {He}\ and\ \citenamefont {Paris}(2025)}]{jadesc}%
  \BibitemOpen
  \bibfield  {author} {\bibinfo {author} {\bibfnamefont {J.}~\bibnamefont
  {He}}\ and\ \bibinfo {author} {\bibfnamefont {M.~G.~A.}\ \bibnamefont
  {Paris}},\ }\href {https://arxiv.org/abs/2503.08235} {\bibinfo {title}
  {Scrambling for precision: optimizing multiparameter qubit estimation in the
  face of sloppiness and incompatibility}} (\bibinfo {year} {2025}),\ \Eprint
  {https://arxiv.org/abs/2503.08235} {arXiv:2503.08235 [quant-ph]} \BibitemShut
  {NoStop}%
\bibitem [{\citenamefont {Hradil}\ \emph {et~al.}(2004)\citenamefont {Hradil},
  \citenamefont {{\v{R}}eh{\'a}{\v{c}}ek}, \citenamefont {Fiur{\'a}{\v{s}}ek},\
  and\ \citenamefont {Je{\v{z}}ek}}]{hradil04}%
  \BibitemOpen
  \bibfield  {author} {\bibinfo {author} {\bibfnamefont {Z.}~\bibnamefont
  {Hradil}}, \bibinfo {author} {\bibfnamefont {J.}~\bibnamefont
  {{\v{R}}eh{\'a}{\v{c}}ek}}, \bibinfo {author} {\bibfnamefont
  {J.}~\bibnamefont {Fiur{\'a}{\v{s}}ek}},\ and\ \bibinfo {author}
  {\bibfnamefont {M.}~\bibnamefont {Je{\v{z}}ek}},\ }\bibinfo {title} {3
  maximum-likelihood methodsin quantum mechanics},\ in\ \href
  {https://doi.org/10.1007/978-3-540-44481-7_3} {\emph {\bibinfo {booktitle}
  {Quantum State Estimation}}},\ \bibinfo {editor} {edited by\ \bibinfo
  {editor} {\bibfnamefont {M.}~\bibnamefont {Paris}}\ and\ \bibinfo {editor}
  {\bibfnamefont {J.}~\bibnamefont {{\v{R}}eh{\'a}{\v{c}}ek}}}\ (\bibinfo
  {publisher} {Springer Berlin Heidelberg},\ \bibinfo {address} {Berlin,
  Heidelberg},\ \bibinfo {year} {2004})\ pp.\ \bibinfo {pages}
  {59--112}\BibitemShut {NoStop}%
\bibitem [{\citenamefont {Mauro~D'Ariano}\ \emph {et~al.}(2000)\citenamefont
  {Mauro~D'Ariano}, \citenamefont {Paris},\ and\ \citenamefont
  {Sacchi}}]{d2000parameter}%
  \BibitemOpen
  \bibfield  {author} {\bibinfo {author} {\bibfnamefont {G.}~\bibnamefont
  {Mauro~D'Ariano}}, \bibinfo {author} {\bibfnamefont {M.~G.~A.}\ \bibnamefont
  {Paris}},\ and\ \bibinfo {author} {\bibfnamefont {M.~F.}\ \bibnamefont
  {Sacchi}},\ }\bibfield  {title} {\bibinfo {title} {Parameter estimation in
  quantum optics},\ }\href {https://doi.org/10.1103/PhysRevA.62.023815}
  {\bibfield  {journal} {\bibinfo  {journal} {Phys. Rev. A}\ }\textbf {\bibinfo
  {volume} {62}},\ \bibinfo {pages} {023815} (\bibinfo {year}
  {2000})}\BibitemShut {NoStop}%
\bibitem [{\citenamefont {Yuen}\ \emph {et~al.}(1975)\citenamefont {Yuen},
  \citenamefont {Kennedy},\ and\ \citenamefont {Lax}}]{Yuen}%
  \BibitemOpen
  \bibfield  {author} {\bibinfo {author} {\bibfnamefont {H.}~\bibnamefont
  {Yuen}}, \bibinfo {author} {\bibfnamefont {R.}~\bibnamefont {Kennedy}},\ and\
  \bibinfo {author} {\bibfnamefont {M.}~\bibnamefont {Lax}},\ }\bibfield
  {title} {\bibinfo {title} {Optimum testing of multiple hypotheses in quantum
  detection theory},\ }\href {https://doi.org/10.1109/TIT.1975.1055351}
  {\bibfield  {journal} {\bibinfo  {journal} {IEEE Transactions on Information
  Theory}\ }\textbf {\bibinfo {volume} {21}},\ \bibinfo {pages} {125} (\bibinfo
  {year} {1975})}\BibitemShut {NoStop}%
\bibitem [{\citenamefont {Paris}(1995)}]{PARIS1995132}%
  \BibitemOpen
  \bibfield  {author} {\bibinfo {author} {\bibfnamefont {M.~G.~A.}\
  \bibnamefont {Paris}},\ }\bibfield  {title} {\bibinfo {title} {Small amount
  of squeezing in high-sensitive realistic interferometry},\ }\href
  {https://doi.org/https://doi.org/10.1016/0375-9601(95)00235-U} {\bibfield
  {journal} {\bibinfo  {journal} {Physics Letters A}\ }\textbf {\bibinfo
  {volume} {201}},\ \bibinfo {pages} {132} (\bibinfo {year}
  {1995})}\BibitemShut {NoStop}%
\bibitem [{\citenamefont {Mukhopadhyay}\ \emph {et~al.}(2025)\citenamefont
  {Mukhopadhyay}, \citenamefont {Bayat}, \citenamefont {Montenegro},\ and\
  \citenamefont {Paris}}]{stepwise}%
  \BibitemOpen
  \bibfield  {author} {\bibinfo {author} {\bibfnamefont {C.}~\bibnamefont
  {Mukhopadhyay}}, \bibinfo {author} {\bibfnamefont {A.}~\bibnamefont {Bayat}},
  \bibinfo {author} {\bibfnamefont {V.}~\bibnamefont {Montenegro}},\ and\
  \bibinfo {author} {\bibfnamefont {M.~G.~A.}\ \bibnamefont {Paris}},\
  }\bibfield  {title} {\bibinfo {title} {Beating joint quantum estimation
  limits with stepwise multiparameter metrology},\ }\bibfield  {journal}
  {\bibinfo  {journal} {arXiv:2506.06075}\ }\href
  {https://doi.org/https://doi.org/10.48550/arXiv.2506.06075}
  {https://doi.org/10.48550/arXiv.2506.06075} (\bibinfo {year}
  {2025})\BibitemShut {NoStop}%
\bibitem [{\citenamefont {Olivares}(2012)}]{olirev2012}%
  \BibitemOpen
  \bibfield  {author} {\bibinfo {author} {\bibfnamefont {S.}~\bibnamefont
  {Olivares}},\ }\bibfield  {title} {\bibinfo {title} {Quantum optics in the
  phase space},\ }\href
  {https://doi.org/https://doi.org/10.1140/epjst/e2012-01532-4} {\bibfield
  {journal} {\bibinfo  {journal} {The European Physical Journal Special
  Topics}\ }\textbf {\bibinfo {volume} {203}},\ \bibinfo {pages} {3} (\bibinfo
  {year} {2012})}\BibitemShut {NoStop}%
\bibitem [{\citenamefont {Olivares}(2021)}]{olirev2021}%
  \BibitemOpen
  \bibfield  {author} {\bibinfo {author} {\bibfnamefont {S.}~\bibnamefont
  {Olivares}},\ }\bibfield  {title} {\bibinfo {title} {Introduction to
  generation, manipulation and characterization of optical quantum states},\
  }\href {https://doi.org/https://doi.org/10.1016/j.physleta.2021.127720}
  {\bibfield  {journal} {\bibinfo  {journal} {Physics Letters A}\ }\textbf
  {\bibinfo {volume} {418}},\ \bibinfo {pages} {127720} (\bibinfo {year}
  {2021})}\BibitemShut {NoStop}%
\end{thebibliography}%

\end{document}